\documentclass[12pt]{article}
\pdfoutput=1
\usepackage{tikz}
\usetikzlibrary{matrix,arrows,decorations.pathmorphing,decorations.pathreplacing}
\usepackage{jheppub}
\usepackage{amsmath,amssymb,euscript,array,mathrsfs,appendix,ctable,marvosym}
\usepackage{arydshln}
\usepackage{todonotes}
\usepackage{graphicx}
\usepackage[normalem]{ulem}

\newcommand\blank[1]{#1}
\renewcommand\blank[1]{}
\def\Buildrel#1\over#2\under#3{\mathrel{\mathop{\kern0pt
#2}\limits^{#1}_{#3}}}

\newcommand{\otoprule}{\midrule[\heavyrulewidth]}

\def\CR{{\Huge\Crossedbox}}
\def\TK{{\Huge\Checkedbox}}
\def\mpsu{\mathfrak{psu}}

\def\mso{\mathfrak {so}}

\def\mh{\mathfrak h}

\def\mR{\mathscr{R}}

\def\msu{\mathfrak{su}}
\def\muu{\mathfrak{u}}

\def\mP{\mathscr{P}}

\def\p{\pi}

\newcommand{\sech}{\operatorname{sech}}
\newcommand{\csch}{\operatorname{csch}}

\def\B0{{\boldsymbol 0}}

\def\CP{{\mathbb C}P}

\def\SU{\text{SU}}

\def\Dbarslash{\,\,{\raise.15ex\hbox{/}\mkern-12mu {\bar D}}}
\def\Dslash{\,\,{\raise.15ex\hbox{/}\mkern-12mu D}}
\def\delslash{\,\,{\raise.15ex\hbox{/}\mkern-9mu \partial}}
\def\delbarslash{\,\,{\raise.15ex\hbox{/}\mkern-9mu {\bar\partial}}}

\def\ket#1{| #1\rangle}

\newcommand{\MAT}[1]{\begin{pmatrix} #1\end{pmatrix}}
\newcommand{\ARR}[1]{\begin{matrix} #1\end{matrix}}
\newcommand{\EQ}[1]{\begin{equation}\begin{split} #1
\end{split}\end{equation}}

\newcommand{\coset}[2]{\raisebox{.5ex}{$#1$}\Big/\raisebox{-.5ex}{$#2$}}
\newcommand{\FIG}[1]{\begin{figure}[ht]\begin{center} #1 \end{center}\end{figure}}
\newcommand{\texpdf}{\texorpdfstring}

\title{
Restoring Unitarity in the $\boldsymbol q$-Deformed World-Sheet S-Matrix}
\author[a]{Ben Hoare,}
\author[b]{Timothy J. Hollowood}
\author[c]{and J. Luis Miramontes}
\affiliation[a]{Institut f\"ur Physik, Humboldt-Universit\"at zu Berlin, Newtonstra\ss e 15, D-12489 Berlin, Germany}
\affiliation[b]{Department of Physics, Swansea University, Swansea, SA2 8PP, U.K.}
\affiliation[c]{Departamento de F\'\i sica de Part\'\i culas and IGFAE,
Universidad
de Santiago de Compostela, 15782 Santiago de Compostela, Spain}

\emailAdd{ben.hoare@physik.hu-berlin.de}
\emailAdd{t.hollowood@swansea.ac.uk}
\emailAdd{jluis.miramontes@usc.es}

\abstract{The world-sheet S-matrix of the string in $\text{AdS}_5\times S^5$ has been shown to admit a $q$-deformation that relates it to the S-matrix of a generalization of the sine-Gordon theory, which arises as the Pohlmeyer reduction of the superstring. Whilst this is a fascinating development the resulting S-matrix is not explicitly unitary. The problem has been known for a long time in the context of S-matrices related to quantum groups. A braiding relation often called ``unitarity" actually only corresponds to quantum field theory unitarity when the S-matrix is Hermitian analytic and quantum group S-matrices manifestly violate this. On the other hand, overall consistency of the S-matrix under the bootstrap requires that the deformation parameter is a root of unity and consequently one is forced to perform the ``vertex" to IRF, or SOS, transformation on the states to truncate the spectrum consistently. In the IRF formulation unitarity is now manifest and the string S-matrix and the S-matrix of the generalised sine-Gordon theory are recovered in two different limits. In the latter case, expanding the Yang-Baxter equation we find that the tree-level S-matrix of the Pohlmeyer-reduced string should satisfy a modified classical Yang-Baxter equation explaining the apparent anomaly in the perturbative computation. We show that the IRF form of the S-matrix meshes perfectly with the bootstrap equations.}

\notoc
\begin{document}

\begin{flushright} \small {\tt HU-EP-13/12}
\end{flushright}
\vspace{0.5cm}

\maketitle

\newpage

\pgfdeclarelayer{top layer}
\pgfdeclarelayer{foreground layer}
\pgfdeclarelayer{background layer}
\pgfsetlayers{background layer,main,foreground layer,top layer}

\section{Introduction}\label{s1}

Integrability has proved to be a very powerful tool in quantum field theory in $1+1$ dimensions. It allows for the exact determination of the S-matrix of a theory based on the symmetries of the underlying QFT. Often the symmetries that arise are modified by the fact that quantum operators have non-trivial exchange properties as a consequence of working in one spatial dimension. Hence in many situations symmetry groups are deformed into quantum groups, which are particular deformations of the universal enveloping algebra of the original Lie algebra.

Of its many applications one of the most striking is to the world-sheet theory of the string moving in space times appearing in the AdS/CFT correspondence. The classic example is the case of the string in AdS$_5 \times S^5$ \cite{Beisert:2010jr,Arutyunov:2009ga}. One of the most practical approaches to quantizing this world-sheet theory has been the use of the BMN light-cone gauge. However, the integrable QFT that arises with this gauge fixing is not relativistically invariant---the energy and momentum do not satisfy the usual relativistic dispersion relation. This lack of relativistic invariance presents considerable complications to the application of the integrable tool box of exact S-matrix theory and the Bethe Ansatz.

We now know that the light-cone gauge-fixed theory, with its coupling constant $g$ (the 't ~Hooft coupling of the dual gauge theory), lies in a larger class of S-matrix theories with a new coupling $k$ \cite{Hoare:2011wr,Hoare:2012fc,Arutyunov:2012zt,Arutyunov:2012ai}, based on the R-matrix associated to the $q$ deformation of the light-cone symmetry algebra \cite{Beisert:2008tw,Beisert:2010kk,Beisert:2011wq,deLeeuw:2011jr}. The string world-sheet theory is obtained in the limit $k\to\infty$ with fixed $g$. Another interesting limit is obtained by taking $g\to\infty$ with fixed $k$, in which case the dispersion relation becomes relativistic and the S-matrix is identified with the S-matrix of a generalized sine-Gordon theory whose classical equation-of-motion is identical to the Pohlmeyer reduction of the equations-of-motion on the string world-sheet \cite{Pohlmeyer:1975nb,Grigoriev:2007bu,Mikhailov:2007xr,Grigoriev:2008jq,Miramontes:2008wt,Hollowood:2009tw,Hollowood:2009sc,Hollowood:2010dt,Hollowood:2011fq}. This generalized sine-Gordon theory involves a WZW model in the bosonic sector and this suggests that $k$ should be a positive integer \cite{Witten:1983ar}. The theories with general $(g,k)$ and their particular limits are illustrated in Figure \ref{f1}. We will argue that a special r\^ole is played by the theories with integer $k$ since these are the ones with a spectrum that naturally truncates. It is very likely that consistent theories only exist with integer $k$.
\FIG{
\begin{tikzpicture} [scale=1.2]
%\draw [help lines] grid (7,5);
\draw[thick] (0,0) -- (0,5) -- (5,5) -- (5,0) -- (0,0);
\node at (-0.4,0) (q1) {$0$};
\node at (-0.4,5) (q2) {$\infty$};
\node at (-0.4,2.5) (q3) {$g$};
%\node at (0,5.3) (q1) {$\infty$};
%\node at (5,5.3) (q2) {$0$};
%\node at (2.5,5.3) (q3) {$g/k$};
\node at (4.5,5.4) (q1) {$3$};
\node at (3.5,5.4) (q2) {$4$};
\node at (0,5.4) (q2) {$\infty$};
\node at (2.5,5.4) (q1) {$5$};
\node at (1.7,5.4) (q2) {$6$};
\node at (1.1,5.4) (q2) {$7$};
%\draw[->] (5.4,2) -- (5.4,4.5);
%\node[rotate=-90] at (5.8,3.25) (a1) {$k$ increasing};
\node at (2.5,5.9) (a1) {$k$};
%\node[rotate=-90] at (6,2.5) (a1) {$g\to\infty$};
\node[rotate=90] at (-2,2.5) (a1) {world-sheet theory in $\text{AdS}_5\times S^5$};
\node[rotate=90] at (-1.5,2.5) (a1) {'t~Hooft coupling $\propto g^2$ ($k\to\infty$)};
\node at (2.5,7) (a1) {generalized sine-Gordon theory};
\node at (2.5,6.5) (a1) {relativistic, SUSY, WZW level $=k$ ($g\to\infty$)};
%\node[rotate=-90] at (6.9,2.7) (a1) {generalized sine-Gordon theory};
%\node[rotate=-90] at (6.4,2.7) (a1) {relativistic, SUSY, WZW level $=k$};
\draw[red] (0.05,5) -- (0.05,0);
%\draw[red] (0.5,5) -- (0.5,0);
%\draw[red] (0.7,5) -- (0.7,0);
\draw[red] (1.1,5) -- (1.1,0);
\draw[red] (1.7,5) -- (1.7,0);
\draw[red] (2.5,5) -- (2.5,0);
\draw[red] (3.5,5) -- (3.5,0);
\draw[red] (4.5,5) -- (4.5,0);
\filldraw[red] (0.3,2.5) circle (0.03cm);
\filldraw[red] (0.55,2.5) circle (0.03cm);
\filldraw[red] (0.8,2.5) circle (0.03cm);
\filldraw[red] (0.3,4.5) circle (0.03cm);
\filldraw[red] (0.55,4.5) circle (0.03cm);
\filldraw[red] (0.8,4.5) circle (0.03cm);
\filldraw[red] (0.3,0.5) circle (0.03cm);
\filldraw[red] (0.55,0.5) circle (0.03cm);
\filldraw[red] (0.8,0.5) circle (0.03cm);
\filldraw[black] (0.05,5) circle (0.05cm);
\filldraw[black] (1.1,5) circle (0.05cm);
\filldraw[black] (1.7,5) circle (0.05cm);
\filldraw[black] (2.5,5) circle (0.05cm);
\filldraw[black] (3.5,5) circle (0.05cm);
\filldraw[black] (4.5,5) circle (0.05cm);
\end{tikzpicture}
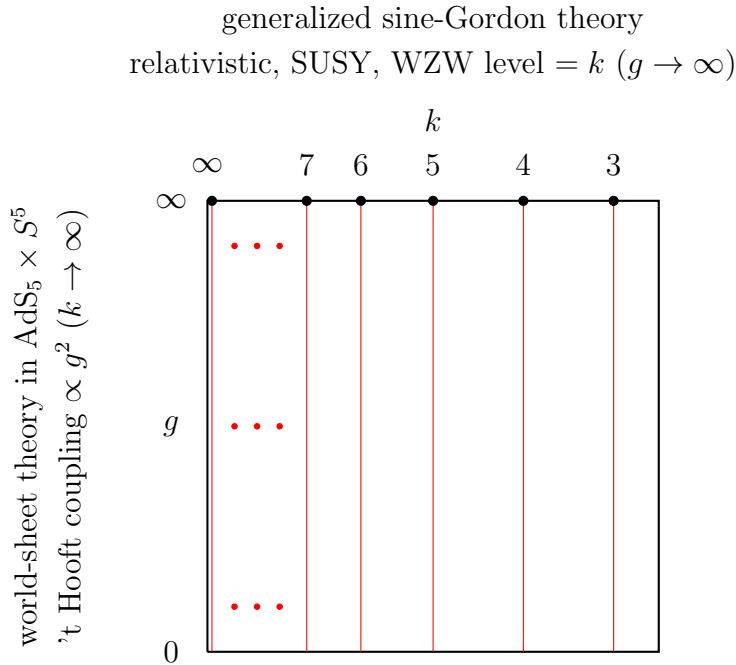
\caption{\small The red lines represent the $q=e^{i\pi/k}$-deformed world-sheet theories for fixed integer $k>2$ and varying $g$. The relativistic generalized sine-Gordon theories are obtained in the limit $g\to\infty$, for fixed integer $k$, while the original world-sheet theory is obtained in the limit $k\to\infty$, fixed $g$.}
\label{f1}
}

A general S-matrix for the basis states of the theory with general $(g,k)$ was written down in \cite{Hoare:2011wr} and initial investigations into its properties have been carried out in \cite{Hoare:2012fc,Arutyunov:2012zt,Arutyunov:2012ai}. It was pointed out in \cite{Arutyunov:2012zt} that the S-matrix theory is not manifestly unitary since its elements are not Hermitian analytic. Ordinarily, Hermitian analyticity, along with the quantum group braiding unitarity, is enough to imply QFT unitarity so a lack of Hermitian analyticity is a serious problem. In \cite{Miramontes:1999gd} it was observed that Hermitian analyticity (and therefore also QFT unitarity) is a basis-dependent identity and hence there is the possibility that there exists a new basis which makes it manifest. On the other hand, there is what seems like a separate issue concerned with how one deals with the representation theory of the quantum group when $k$ is an integer and the deformation parameter $q=e^{i\pi/k}$ is a root of unity. The latter issue is well known in the context of integrable lattice models and requires changing from the vertex picture where the states, in the case of an $\SU(2)$ symmetry, transform in the doublet, to the Interaction-Round-a-Face (IRF), also know as the Solid-On-Solid (SOS), picture, where states correspond to kinks between vacua labelled by highest weight states, that is arbitrary spins: see Figure \ref{f2}.
\FIG{
\begin{tikzpicture}[scale=0.8]
\begin{scope}[scale=0.8]
\filldraw[black] (0,0) circle (4pt);
    \draw (-1,-1) -- (1,1);
    \draw (-1,1) -- (1,-1);
    \node at (-1.5,-1.5) (a1) {$i$};
       \node at (1.5,-1.5) (a1) {$j$};
   \node at (-1.5,1.5) (a1) {$k$};
      \node at (1.5,1.5) (a1) {$l$};
\draw[double,<->] (2.5,0) -- (4.5,0);
\end{scope}
  \begin{scope}[xshift=6cm,scale=0.8]
      \draw (-1,-1) -- (-1,1) -- (1,1) -- (1,-1) -- (-1,-1);
         \node at (-1.5,0) (a1) {$a$};
       \node at (0,-1.5) (a2) {$b$};
           \node at (1.5,0) (a3) {$c$};
    \node at (0,1.5) (a4) {$d$};
    \end{scope}
\end{tikzpicture}
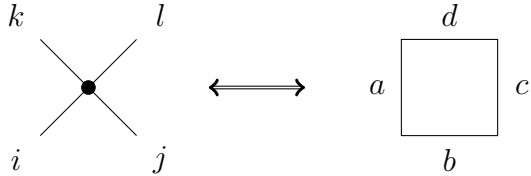
\caption{\small The vertex and IRF labels for the Boltzmann weights of an integrable lattice model. In the vertex picture the labels $i,j,k,l\in\{\pm\frac12\}$ are the weights of the spin $\frac12$ representation of $\SU(2)$. In the IRF picture the labels $a,b,c,d\in\{0,\frac12,1,\frac32,\ldots\}$ are spins of arbitrary irreducible representations of $\SU(2)$ with $|a-b|=|b-c|=|a-d|=|d-c|=\frac12$.}
\label{f2}
}
This vertex-to-IRF transformation was used in the context of S-matrix theory by Bernard and LeClair in the example of the sine-Gordon soliton S-matrix in \cite{LeClair:1989wy,Bernard:1990cw,Bernard:1990ys}. What is remarkable, and as far as we know unrecognised in the old quantum group S-matrix literature, is that the IRF form of the S-matrix is manifestly Hermitian analytic and consequently satisfies QFT unitarity.\footnote{Some relevant references for this old literature are
\cite{deVega:1990av,Hollowood:1992sy,Hollowood:1993fj,Delius:1995he,Gandenberger:1997pk,MacKay:1990mp}.} Whilst the $\SU(2)$ example of Bernard and LeClair is simple enough that the original vertex form of the S-matrix can be rendered Hermitian analytic by a suitable conjugation, this does not work for $\SU(N)$ generalizations and was an outstanding puzzle. Bernard and LeClair went on to argue that the IRF, or SOS, form for the S-matrix was the ideal basis for dealing with the complications of the representation theory of the $U_q(\msu(2))$ quantum group when $q$ is a root of unity which are summarized in Appendix~\ref{a5}. In fact, the IRF form of the S-matrix naturally projects out the ``bad" representations in the Hilbert space to leave a perfectly consistent S-matrix. In statistical mechanics language this restriction leads to the Restricted-Solid-On-Solid RSOS lattice models. In this work we will argue that the vertex-to-IRF transformation is exactly what is needed to make QFT unitarity manifest for the general $(g,k)$ theories and, at the same time, the IRF picture is perfectly adapted to implementing the RSOS-like restriction on the Hilbert space when $k$ is an integer. These reduced theories are likely to be the only consistent S-matrix theories with an finite spectrum.

Further evidence that the kink picture is the most natural basis comes from studying various limits of the S-matrix in the IRF picture. Firstly, we find that if we take the limit in which $k \rightarrow \infty$ (up to some minor subtleties) we recover the by now well-known light-cone string S-matrix \cite{Beisert:2005tm,Klose:2006zd}. This should be expected as this corresponds to the limit in which $q$, the quantum deformation parameter, goes to unity. More unexpected is what happens when we take the limit $g \rightarrow \infty$, which is meant to correspond to the Pohlmeyer-reduced theory. Expanding around large $k$ (again up to some subtleties) we find that the tree-level S-matrix should satisfy a modified classical Yang-Baxter equation. Furthermore this term in the expansion agrees with the perturbative computation of \cite{Hoare:2009fs}, explaining the apparent anomaly in this calculation.

The kink picture also arises naturally in the context of the Pohlmeyer-reduced theory, which is a generalized sine-Gordon theory whose Lagrangian action includes a WZW term in the bosonic sector. Its definition requires careful treatment due to the existence of field configurations (solitons) with non-trivial boundary conditions~\cite{Hollowood:2011fq}. Namely, it has to be defined on a world-sheet with boundary, and its consistency imposes quantization conditions on the boundary conditions themselves~\cite{Alekseev:1998mc,Gawedzki:2001ye,Elitzur:2001qd}, in addition to the well known quantization of the coupling constant~\cite{Witten:1983ar}. 
The resulting picture is that soliton solutions are kinks that interpolate between a finite set of vacua labelled by highest weight states~\cite{Hollowood:2013oca}.

Before we proceed it is worthwhile stating the S-matrix axioms of relevance in the context of a non-relativistic theory. Due to integrability an $n$-body S-matrix element factorizes into 2-body building blocks for which the separate rapidities are preserved. However the S-matrix is a function of the rapidities separately $S(\theta_1,\theta_2)$ and not just of $\theta_1-\theta_2$ as in a relativistic theory. The rapidity is a familiar variable in $1+1$-dimensions determining the velocity via $v=\tanh\theta$, however the relativistic relations $E=m\cosh\theta$ and $p=m\sinh\theta$ are {\em not\/} satisfied.

The S-matrix elements generally satisfy two important identities that are known as crossing symmetry and ``braiding unitarity". It is important that the latter is not the same as ``QFT unitarity", which is the familiar requirement that the S-matrix is a unitary matrix for physical (real) rapidities. Assuming that particles come in multiplets whose states are labelled by $i$, crossing symmetry implies
\EQ{
S_{ij}^{kl}(\theta_1,\theta_2)={\cal C}_{kk'}S_{k'i}^{lj'}(i\pi+\theta_2,\theta_1){\cal C}^{-1}_{j'j}\ ,
\label{cross}
}
where ${\cal C}$ is the charge conjugation matrix. The braiding unitarity relation takes the form
\EQ{
\sum_{kl}S_{ij}^{kl}(\theta_1,\theta_2)S_{kl}^{mn}(\theta_2,\theta_1)=\delta_{im}\delta_{jn}\ ;
\label{bru}
}
however, this only implies QFT unitarity, that is 
\EQ{
\sum_{kl}S_{ij}^{kl}(\theta_1,\theta_2)S_{mn}^{kl}(\theta_1,\theta_2)^*=\delta_{im}\delta_{jn}\ ,\qquad (\theta_i\text{ real})\ ,
\label{qft}
}
when the S-matrix satisfies Hermitian analyticity:
\EQ{
S_{ij}^{kl}(\theta_1^*,\theta_2^*)^*=S_{kl}^{ij}(\theta_2,\theta_1)\ .
\label{ha}
}

The other S-matrix axioms govern the structure of bound states and the bootstrap equations. For completeness we have included a discussion of the bootstrap programme in Appendix \ref{a4}. Overall, non-relativistic integrable S-matrix theory enjoys most of the properties of relativistic integrable S-matrix theory as we summarize in the check-list below:
\begin{center}
\begin{tabular}{lcc}
\toprule
&Relativistic & Non-relativistic  \\
\otoprule
Factorization & \TK &\TK\\
Function of rapidity difference & \TK & \CR\\
Meromorphic function of rapidity & \TK & \CR\\
Hermitian analyticity & \TK & \TK\\
QFT Unitarity & \TK &\TK\\
Crossing & \TK & \TK\\
Bound-state poles &\TK&\TK\\
Bootstrap&\TK&\TK\\
\bottomrule
\end{tabular}
\end{center}
In the non-relativistic case the only difference is that S-matrix elements like $S(\theta_1,\theta_2)$ are not functions of the rapidity difference since there is no boost invariance  and the S-matrix is not a meromorphic function on the complex rapidity plane; on the contrary, there are branch cuts~\cite{Hoare:2011wr}. However, all other properties hold just as in the relativistic case.

\section{Lessons from the Restricted Sine-Gordon Theory}\label{s2}

S-matrices for relativistic integrable quantum field theories are built out of solutions to the Yang-Baxter equation, for which quantum groups provide an algebraic framework. The simplest solution is related to the quantum group deformation of the affine (loop) Lie algebra $\msu(2)^{(1)}$. The basis states $\ket{\phi_m}$ transform in the spin $\frac12$ representation with $m=\pm\frac12$. If $V_{j}$ is the spin $j$ representation space then the two-body S-matrix, from which the complete S-matrix is constructed by factorization, is a map (known as an intertwiner)
\EQ{
S(\theta):\qquad V_{\frac12}(\theta_1)\otimes V_{\frac12}(\theta_2)\longrightarrow V_{\frac12}(\theta_2)\otimes V_{\frac12}(\theta_1)\ .
}
Here, we have indicated the rapidity of the states by $\theta_i$, and $\theta=\theta_1-\theta_2$. The S-matrix takes the form
\EQ{
S(\theta)=v(\theta)\check R(x(\theta))\ ,
}
where $x(\theta)=e^{\lambda\theta}$, $\check R(x)$ is the ``$R$-matrix" of the affine quantum group and $v(\theta)$ is a scalar factor, in current parlance the ``dressing phase", required to ensure all the necessary S-matrix axioms are satisfied. To make connection with~\eqref{cross}--\eqref{ha}, we define the S-matrix elements so that
\EQ{
\ket{\phi_i(\theta_1)}\otimes\ket{\phi_j(\theta_2)}\longrightarrow S_{ij}^{kl}(\theta)\, \ket{\phi_k(\theta_2)}\otimes \ket{\phi_l(\theta_1)}\,.
}

The non-trivial elements of the S-matrix are
\EQ{
\ket{\phi_{\pm\frac12}(\theta_1)\phi_{\pm\frac12}(\theta_2)}&\longrightarrow v(\theta)(qx-q^{-1}x^{-1})\,\ket{\phi_{\pm\frac12}(\theta_2)\phi_{\pm\frac12}(\theta_1)}\ ,\\
\ket{\phi_{\pm\frac12}(\theta_1)\phi_{\mp\frac12}(\theta_2)}&\longrightarrow v(\theta)(x-x^{-1})\,\ket{\phi_{\mp\frac12}(\theta_2)\phi_{\pm\frac12}(\theta_1)}\\ &~~~~~~~~~~~~~~~~~+
v(\theta)x^{\pm1}(q-q^{-1})\,\ket{\phi_{\pm\frac12}(\theta_2)\phi_{\mp\frac12}(\theta_1)}\ .
\label{wea}
}
There are consequently three basic processes; ``identical particle", ``transmission" and ``reflection" (although two of the latter):
\EQ{
S_I(\theta)&=\begin{tikzpicture}[baseline=-0.65ex,scale=0.4]
\filldraw[black] (0,0) circle (4pt);
    \draw[->] (-0.6,-0.6) -- (0.6,0.6);
    \draw[<-] (-0.6,0.6) -- (0.6,-0.6);
    \node at (-1.3,1.3) (a1) {$\pm\frac12$};
       \node at (1.3,-1.3) (a2) {$\pm\frac12$};
           \node at (1.3,1.3) (a3) {$\pm\frac12$};
    \node at (-1.3,-1.3) (a4) {$\pm\frac12$};
\end{tikzpicture}(\theta)=v(\theta)(xq-q^{-1}x^{-1})\ ,\\
S_T(\theta)&=\begin{tikzpicture}[baseline=-0.65ex,scale=0.5]
\filldraw[black] (0,0) circle (4pt);
    \draw[->] (-0.6,-0.6) -- (0.6,0.6);
    \draw[<-] (-0.6,0.6) -- (0.6,-0.6);
    \node at (-1.3,1.3) (a1) {$\mp\frac12$};
       \node at (1.3,-1.3) (a2) {$\mp\frac12$};
           \node at (1.3,1.3) (a3) {$\pm\frac12$};
    \node at (-1.3,-1.3) (a4) {$\pm\frac12$};
\end{tikzpicture}(\theta)=v(\theta)(x-x^{-1})\ ,\\
S_R^\pm(\theta)&=\begin{tikzpicture}[baseline=-0.65ex,scale=0.5]
\filldraw[black] (0,0) circle (4pt);
    \draw[->] (-0.6,-0.6) -- (0.6,0.6);
    \draw[<-] (-0.6,0.6) -- (0.6,-0.6);
    \node at (-1.3,1.3) (a1) {$\pm\frac12$};
       \node at (1.3,-1.3) (a2) {$\mp\frac12$};
           \node at (1.3,1.3) (a3) {$\mp\frac12$};
    \node at (-1.3,-1.3) (a4) {$\pm\frac12$};
\end{tikzpicture}(\theta)=v(\theta)x^{\pm1}(q-q^{-1})\ .
\label{wea2}
}

There are two theories whose S-matrices are built out of the $R$-matrix of the $\msu(2)$ quantum group. The first is associated to solitons of the sine-Gordon theory for which \cite{Jimbo:1985zk,Jimbo:1985vd,Jimbo:1987ra,Jimbo:1989qm,Bernard:1990cw,Bernard:1990ys}
\EQ{
x=e^{\frac{\theta}{k}}\ ,\qquad q=-e^{-i\pi/k}\ ,
\label{de1}
}
and the second to certain symmetric space sine-Gordon (SSSG) theories. In particular, we have in mind
the $S^5$ SSSG theory whose underlying quantum group symmetry
is $\mso(4) \cong \msu(2) \oplus \msu(2)$ and, accordingly, the S-matrix should factorize into two copies of the $R$-matrix of the $\msu(2)$ quantum group up to an overall phase.\footnote{The same $R$-matrix also describes the  ${\mathbb C}P^3$ SSSG theory whose underlying symmetry is $\muu(2) \cong \msu(2) \oplus\muu(1)$\cite{Hollowood:2010rv}.} In this case, we have
\EQ{
x=e^{\frac{k+1}{k+2}\theta}\ ,\qquad q=e^{\frac{i\pi}{k+2}}\ .
\label{de2}
}
In both cases \eqref{de1} and \eqref{de2}
\EQ{
x(\theta)=-q^{-1}x(i\pi-\theta)^{-1}\ ,
}
which implies
\EQ{
S_I(\theta)=S_T(i\pi-\theta)\ ,\qquad S_R^+(\theta)=-q^{-1}S_R^-(i\pi-\theta)\ ,
}
as long as the dressing phase satisfies $v(\theta)=v(i\pi-\theta)$.
Crossing symmetry is then ensured by defining charge conjugation as
\EQ{
{\cal C}\ket{\phi_{\pm\frac12}(\theta)}=\pm i q^{\pm1/2}\ket{\phi_{\mp\frac12}(\theta)}\ .
\label{phicross}
}

The S-matrix satisfies the braiding relation
\EQ{
\sum_{kl}S_{ij}^{kl}(\theta)S_{kl}^{mn}(-\theta)=v(\theta)v(-\theta)(qx-q^{-1}x^{-1})(qx^{-1}-q^{-1}x)\delta_{im}\delta_{jn}\ .
}
Therefore, as long as the dressing phase satisfies
\EQ{
v(\theta)v(-\theta)=\frac1{(qx-q^{-1}x^{-1})(qx^{-1}-q^{-1}x)}\ ,
\label{phasecross}}
the S-matrix satisfies the braiding unitarity relation \eqref{bru}. However, this is not equivalent to QFT unitarity because as it stands the S-matrix written in this ``particle''-like basis does not satisfy Hermitian analyticity \eqref{ha} \cite{Miramontes:1999gd}. Whilst, given that the dressing factor satisfies $v(\theta^*)^*=-v(-\theta)$, the identical particle and transmission elements satisfy the required identity
\EQ{
S_I(\theta^*)^*=S_I(-\theta)\ ,\qquad S_T(\theta^*)^*=S_T(-\theta)\ ,
}
the reflection amplitudes are non-compliant because they satisfy
\EQ{
S_R^\pm(\theta^*)^*=S_R^\mp(-\theta)\ ,
}
rather than $S_R^\pm(\theta^*)^*=S_R^\pm(-\theta)$,
clearly violating \eqref{ha}.

In the case of $\msu(2)$, Hermitian analyticity can be restored by a simple rapidity-dependent transformation on the states of the form \cite{Jimbo:1985zk,Jimbo:1985vd,Jimbo:1987ra,Jimbo:1989qm,Bernard:1990cw,Bernard:1990ys}
\EQ{
\ket{\phi_{\pm\frac12}(\theta)}\longrightarrow x(\theta)^{\pm1/2}\ket{\phi_{\pm\frac12}(\theta)}\ .
}
This transformation removes the factors of $x^{\pm1}$ from the reflection amplitudes and restores Hermitian analyticity.\footnote{To ensure crossing symmetry charge conjugation needs to be modified so that ${\cal C} \ket{\phi_{\pm \frac12}} = \ket{\phi_{\mp \frac12}}$, in agreement with the original construction of \cite{Zamolodchikov:1978xm,Dorey:1996gd}
.}
It has an algebraic interpretation of moving from the homogeneous to the principal gradation of the affine algebra $\msu(2)^{(1)}$.
The new S-matrix is precisely the S-matrix of the solitons of the sine-Gordon theory. In is interesting to note that the same kind of transformation is not sufficient to restore Hermitian analyticity for the $\msu(n)$ generalization of the S-matrix and this deficiency of quantum group S-matrices was never resolved in the literature.

However, there is another way to restore Hermitian analyticity that does not involve changing the gradation and can be generalized to higher rank algebras \cite{Ahn:1990gn,Hollowood:1992sy}. This is the vertex-to-IRF transformation \cite{LeClair:1989wy,Bernard:1990cw,Bernard:1990ys}. 
In the context of the sine-Gordon theory, the 
transformation is a mathematical procedure that takes the S-matrix of one theory (the sine-Gordon theory) and produces the S-matrix of a new one (the restricted sine-Gordon theory). At this stage it is worth recalling that the 1-particle states in the sine-Gordon theory are labelled by $m=\pm\frac12$, which is a $U(1)$ topological charge, while in the restricted sine-Gordon theory the 1-particle states are labelled by a pair of spins $(j,j')$ with $|j-j'|=\frac12$ and $j,j'\in\{0,\frac12,\ldots,j_{\text{max}}\}$. 

The vertex-to-IRF transformation involves two steps. The first one is simply a change of basis in the Hilbert space of multi-particle states 
$\ket{\phi_{m_1}(\theta_1)\phi_{m_2}(\theta_2)\cdots\phi_{m_N}(\theta_N)}$ which transform in the tensor product representation ${V_{\frac12}}^{\otimes N}$ of the quantum group.
This is the ``vertex" basis. The new basis corresponds to decomposing the multi-particle states into irreducible representations. The group theory here is analogous to the decomposition of representations of $\msu(2)$ (at least when $q$ is generic). In the new basis, the $N$-soliton states that transform in the representation of total spin $J$, say $\ket{{\cal J},M}$, are labelled by the chain of the spins in the decomposition, ${\cal J}=(j_1\equiv J,j_2,\ldots,j_N\equiv\frac12)$ with $|j_{i-1}-j_i|=\frac12$, and by the $j_z$ quantum number $-J\leq M\leq+J$. In order to describe the change of basis we will need the  $q$-deformed Clebsch-Gordan ($q$-CG) coefficients which we define by
\EQ{
\ket{J,M}=\sum_{m_1,m_2}\left[\ARR{J & j_1 & j_2\\
M & m_1 & m_2}\right]_q\ket{j_1,m_1}\otimes\ket{j_2,m_2}\ ,
\label{qCGor}
}
where the sum is taken with $M=m_1+m_2$ fixed. The $q$-CG coefficients that we need are given in Appendix \ref{A1}.
In terms of them, the change of basis takes the form
\EQ{
\ket{{\cal J},M;\{\theta_i\}}&=\sum_{\{m_i=\pm\frac12\}}\left[\ARR{j_1 & \tfrac12 & j_2\\
M & m_1 & M_2}\right]_q
\left[\ARR{j_2 & \frac12 & j_3\\
M_2 & m_2 & M_3}\right]_q\cdots\\[5pt]
 &\hspace{1.5cm}
\cdots
\left[\ARR{j_{N-1} & \tfrac12 & \frac12\\
M_{N-1} & m_{N-1} & m_N}\right]_q \ket{\phi_{m_1}(\theta_1)\phi_{m_2}(\theta_2)\cdots\phi_{m_N}(\theta_N)}\ ,
\label{VtI}
}
where $M_{i}=M_{i+1}+m_i$, $M_1\equiv M$, $M_N\equiv m_N$, and the sum is over all the $\{m_i=\pm\frac12\}$ subject to $M=\sum_{i=1}^Nm_i$ being fixed.
The new basis can be interpreted in terms of the states
\EQ{
\ket{\Phi_{j_1j_2}(\theta)}^{M_1}{}_{M_2}=\sum_{m=\pm\frac12}\left[\ARR{j_1 & \tfrac12 & j_2\\
M_1 & m & M_2}\right]_q\ket{\phi_m (\theta)}\,,
\label{newop}
}
so that
\EQ{
\ket{(j_1,j_2,\ldots,j_N),M_1;\{\theta_i\}}&=
\ket{\Phi_{j_1j_2}(\theta_1)\Phi_{j_2j_3}(\theta_2)\cdots \Phi_{j_N0}(\theta_N)}^{M_1}{}_0\,,
\label{mpsL}
}
which involves an implicit sum over $M_2,M_3,\ldots, M_{N}$, and where the product satisfies the adjacency conditions $|j_i-j_{i+1}|=j_N=\frac12$.

A crucial part of the vertex-to-IRF transformation is the observation that, in this basis, the two-body S-matrix elements are given by
\EQ{
&\ket{\Phi_{j_{i-1}\, j_i}(\theta_{i-1})\,\Phi_{j_{i}\, j_{i+1}}(\theta_{i})}^{M_{i-1}}{}_{M_{i+1}} \\[5pt]
&
\qquad\qquad\longrightarrow
\sum_{j_i'}\begin{tikzpicture}[baseline=-0.65ex,scale=0.6]
    \draw (-0.6,-0.6) -- (-0.6,0.6) -- (0.6,0.6) -- (0.6,-0.6) -- (-0.6,-0.6);
    \node at (-1.4,0) (a1) {$j_{i-1}$};
       \node at (0,-1.1) (a2) {$j_i$};
           \node at (1.4,0) (a3) {$j_{i+1}$};
    \node at (0,1.2) (a4) {$j_i'$};
\end{tikzpicture}(\theta)\,\,
\ket{\Phi_{j_{i-1}\, j'_i}(\theta_{i})\Phi_{j'_{i}\, j_{i+1}}(\theta_{i-1})}^{M_{i-1}}{}_{M_{i+1}} \,,
\label{smatL}
}
which is illustrated in Appendix \ref{A1}. Here, the box denotes a function of the rapidity difference $\theta=\theta_{i-1}-\theta_i$ and of the spins $j_{i-1}$, $j_i$, $j_{i+1}$ and $j_i'$ given by~\eqref{gtw}.
In other words, the two-body S-matrix elements are diagonal in the $j_z$ quantum numbers and completely independent of them.
This  fact is a consequence of the quantum group invariance of the $R$-matrix, but at an explicit level it will be important later to notice that it depends on the specific identity
\EQ{
S_R^\pm(x)
+q^{\mp1}S_T(x)=S_I(x)\ ,
\label{kid}
}
that is
\EQ{
x^{\pm1}(q-q^{-1})+q^{\mp1}(x-x^{-1})=qx-q^{-1}x^{-1}\ .
\label{sid}
}
Eq.~\eqref{smatL} motivates the second step of the vertex-to-IRF transformation, which simply amounts to
being blind to the $j_z$ quantum numbers
\EQ{
\ket{\Phi_{j_1\, j_2}(\theta)}^{M_1}{}_{M_2} \longrightarrow \ket{K_{j_1\, j_2}(\theta)}\,.
}
The multi-particle states~\eqref{mpsL} corresponding to different values of $M_1$ are then identified with a single state
\EQ{
\ket{K_{j_1j_2}(\theta_1)K_{j_2j_3}(\theta_2)\cdots K_{j_N 0}(\theta_N)}\,.
\label{mkL}
}
This is the tensor product of $N$ one-particle states $\ket{K_{j j'}(\theta)}$ labelled by two spins with $|j-j'|=\frac12$ which are scalar with respect to the quantum group. 
They can be naturally interpreted as describing a set of kinks in a theory with a degenerated set of vacua labelled by $j\in\{0,\tfrac12,1,\tfrac32,\ldots\}$ and, thus, associated to the highest weight representations of $U_q(\msu(2))$. Within this interpretation, $\ket{K_{j j'}(\theta)}$ corresponds to a kink between neighboring vacua, and~\eqref{mkL} to an $N$-kink state 
between the vacua labelled by 0 and $j_1$. 
In this IRF picture, it is natural to generalize the multi-kink states~\eqref{mkL} by allowing them to be built on an arbitrary vacuum on the right. The more general states are then
\EQ{
\ket{K_{j_1j_2}(\theta_1)K_{j_2j_3}(\theta_2)\cdots K_{j_N j_{N+1}}(\theta_N)}\,,
\label{mkL2}
}
which is an $N$-kink state 
between the vacua labelled by $j_{N+1}$ and $j_1$.

In the context of~\eqref{VtI} and~\eqref{mpsL}, the vacuum labelled by $j_1$ has a 
moduli space corresponding to the whole irreducible representation $V_{j_1}$, with basis $\ket{j_1,M_1}$ for $-j_1\leq M_1\leq +j_1$. 
In a QFT in $1+1$ dimensions a continuous symmetry cannot be spontaneously broken and operationally this means that potential vacuum moduli spaces should actually be integrated over in the functional integral. In the present situation, since S-matrix elements do not depend on the points in the moduli spaces and, in particular, on the quantum number $M_1$, performing the integrals simply amounts to being blind to the quantum number $M_1$, which  makes the vertex-to-IRF transformation natural.

The vertex-to-IRF transformation is particularly relevant in the case when $k$ is a positive integer and, hence,
$q$ is a root of unity. Then, the number of irreducible representations $V_j$ is bounded by $k$ and, moreover, the tensor product of two of them decomposes as a sum of both irreducible and  {reducible but indecomposable} representations.\footnote{The main features of the representation theory of $U_q(\msu(2))$ are summarized in Appendix~\ref{a5}.}
For $q=e^{i\pi/k}$, the tensor product can be consistently restricted to the irreducible representations $V_j$ with $j=0,\tfrac12,1,\ldots, \tfrac{k}2-1$, so that it becomes~\eqref{ttp}. Correspondingly,
the S-matrix theory naturally preserves the sub-sector of states formed from kinks associated to the vacua labelled by the subset $j\in\{0,\frac12,\ldots,\frac{k}2-1\}$. This restriction corresponds to taking the restricted-SOS, or RSOS, models of statistical physics. In the context of the sine-Gordon theory the restricted model is known as the restricted $k/(k+1)$ sine-Gordon theory which involves $2(k-2)$ different elementary kinks.

According to~\eqref{smatL}, the two-body S-matrix elements in the IRF picture correspond to the pairwise processes
\EQ{
\ket{\cdots K_{j_{i-1}j_i}(\theta_{i-1})K_{j_ij_{i+1}}(\theta_i)\cdots}\longrightarrow
\ket{\cdots K_{j_{i-1}j'_i}(\theta_i)K_{j'_ij_{i+1}}(\theta_{i-1})\cdots}\ .
}
As we show in Appendix \ref{A1}, the explicit non-trivial elements of the S-matrix are
\EQ{
&\ket{K_{j\pm1,j\pm\frac12}(\theta_1)K_{j\pm\frac12,j}(\theta_2)}\longrightarrow S_I(\theta)\ket{K_{j\pm1,j\pm\frac12}(\theta_2)K_{j\pm\frac12,j}(\theta_1)}\ ,\\[5pt]
&\ket{K_{j,j\pm\frac12}(\theta_1)K_{j\pm\frac12,j}(\theta_2)}\longrightarrow\frac{\sqrt{[2j][2j+2]}}{[2j+1]}S_T(\theta)\ket{K_{j,j\mp\frac12}(\theta_1)K_{j\mp\frac12,j}(\theta_2)}\\[5pt] 
&\hspace{131pt}+\frac{q^{2j+1}S_R^\mp(\theta)-q^{-2j-1}S_R^\pm(\theta)}{q^{2j+1}-q^{-2j-1}}
\ket{K_{j,j\pm\frac12}(\theta_1)K_{j\pm\frac12,j}(\theta_2)}\ .
\label{web}
}
In this formula, we have used the $q$-number
\EQ{
[n]=\frac{q^n-q^{-n}}{q-q^{-1}}\ .
\label{qnb}
}
and $\theta=\theta_1-\theta_2$.
We can summarize the complete S-matrix in the following way:\EQ{
\begin{tikzpicture}[baseline=-0.65ex,scale=0.6]
    \draw (-0.6,-0.6) -- (-0.6,0.6) -- (0.6,0.6) -- (0.6,-0.6) -- (-0.6,-0.6);
    \node at (-1,0) (a1) {$a$};
       \node at (0,-1) (a2) {$b$};
           \node at (1,0) (a3) {$c$};
    \node at (0,1) (a4) {$d$};
\end{tikzpicture}(\theta)
:\quad K_{ab}(\theta_1)+K_{bc}(\theta_2)\longrightarrow K_{ad}(\theta_2)+K_{dc}(\theta_1)\ ,
}
where the kinks $K_{ab}(\theta)$ are required to satisfy the adjacency condition $|a-b|=\frac12$. The S-matrix can then be written compactly as
\EQ{
\begin{tikzpicture}[baseline=-0.65ex,scale=0.6]
    \draw (-0.6,-0.6) -- (-0.6,0.6) -- (0.6,0.6) -- (0.6,-0.6) -- (-0.6,-0.6);
    \node at (-1,0) (a1) {$a$};
       \node at (0,-1) (a2) {$b$};
           \node at (1,0) (a3) {$c$};
    \node at (0,1) (a4) {$d$};
\end{tikzpicture}(\theta)=S_I(\theta)\delta_{bd}+e^{i\pi(a+c-b-d)}S_T(\theta)\sqrt{\frac{[2b+1][2d+1]}{[2a+1][2c+1]}}\, \delta_{ac}\ .
\label{gtw}
}
Using this notation, it is worth noticing that the independence of the S-matrix element~\eqref{smatL} on the $j_z$ quantum numbers corresponds to the identity
\EQ{
&\sum_{\alpha,\beta} \left[\ARR{j_{i-1} & \tfrac12 & j_i\\
M_{i-1} & \alpha & M_i}\right]_q 
\left[\ARR{j_i & \tfrac12 & j_{i+1}\\
M_i & \beta & M_{i+1}}\right]_q \, S_{\alpha\beta}^{\beta'\alpha'}(\theta)\\[5pt]
&\qquad
=\sum_{j'_i} 
\begin{tikzpicture}[baseline=-0.65ex,scale=0.6]
    \draw (-0.6,-0.6) -- (-0.6,0.6) -- (0.6,0.6) -- (0.6,-0.6) -- (-0.6,-0.6);
    \node at (-1.3,0) (a1) {$j_{i-1}$};
       \node at (0,-1) (a2) {$j_i$};
           \node at (1.3,0) (a3) {$j_{i+1}$};
    \node at (0,1) (a4) {$j'_i$};
\end{tikzpicture}(\theta)
\left[\ARR{j_{i-1} & \tfrac12 & j'_i\\
M_{i-1} & \beta' & M_i'}\right]_q 
\left[\ARR{j'_i & \tfrac12 & j_{i+1}\\
M_i' & \alpha' & M_{i+1}}\right]_q\ .
}

\noindent{\bf Crossing symmetry}
\nopagebreak

The S-matrix satisfies the following crossing symmetry relation
\EQ{
\begin{tikzpicture}[baseline=-0.65ex,scale=0.6]
    \draw (-0.6,-0.6) -- (-0.6,0.6) -- (0.6,0.6) -- (0.6,-0.6) -- (-0.6,-0.6);
    \node at (-1,0) (a1) {$a$};
       \node at (0,-1) (a2) {$b$};
           \node at (1,0) (a3) {$c$};
    \node at (0,1) (a4) {$d$};
\end{tikzpicture}(\theta)=e^{i\pi(a+c-b-d)}\sqrt{\frac{[2b+1][2d+1]}{[2a+1][2c+1]}}\begin{tikzpicture}[baseline=-0.65ex,scale=0.6]
    \draw (-0.6,-0.6) -- (-0.6,0.6) -- (0.6,0.6) -- (0.6,-0.6) -- (-0.6,-0.6);
    \node at (-1,0) (a1) {$d$};
       \node at (0,-1) (a2) {$a$};
           \node at (1,0) (a3) {$b$};
    \node at (0,1) (a4) {$c$};
\end{tikzpicture}(i\pi-\theta)\ ,
}
which is the correct crossing equation with charge conjugation defined as
\EQ{
{\cal C}\ket{K_{ab}(\theta)}=e^{i\pi(b-a)}\sqrt{\frac{[2a+1]}{[2b+1]}}\, \ket{K_{ba}(\theta)} \ .
}
This is consistent with the charge conjugation of $\ket{\phi_{\pm\frac12}(\theta)}$ \eqref{phicross} and the vertex-to-IRF transformation \eqref{VtI} as shown in Appendix \ref{A1}.

\noindent{\bf Hermitian analyticity}
\nopagebreak

The resulting S-matrix now satisfies the kink version of the Hermitian analyticity
\EQ{
\begin{tikzpicture}[baseline=-0.65ex,scale=0.6]
    \draw (-0.6,-0.6) -- (-0.6,0.6) -- (0.6,0.6) -- (0.6,-0.6) -- (-0.6,-0.6);
    \node at (-1,0) (a1) {$a$};
       \node at (0,-1) (a2) {$b$};
           \node at (1,0) (a3) {$c$};
    \node at (0,1) (a4) {$d$};
\end{tikzpicture}(\theta^*)^*=
\begin{tikzpicture}[baseline=-0.65ex,scale=0.6]
    \draw (-0.6,-0.6) -- (-0.6,0.6) -- (0.6,0.6) -- (0.6,-0.6) -- (-0.6,-0.6);
    \node at (-1,0) (a1) {$a$};
       \node at (0,-1) (a2) {$d$};
           \node at (1,0) (a3) {$c$};
    \node at (0,1) (a4) {$b$};
\end{tikzpicture}(-\theta)\ .
\label{kha}
}
Note that it is important that in each case the expressions inside the square roots in \eqref{web} and \eqref{gtw} are real and positive. In the kink basis the braiding unitarity relation takes the form
\EQ{
\sum_e
\begin{tikzpicture}[baseline=-0.65ex,scale=0.6]
    \draw (-0.6,-0.6) -- (-0.6,0.6) -- (0.6,0.6) -- (0.6,-0.6) -- (-0.6,-0.6);
    \node at (-1,0) (a1) {$a$};
       \node at (0,-1) (a2) {$b$};
           \node at (1,0) (a3) {$c$};
    \node at (0,1) (a4) {$e$};
\end{tikzpicture}(\theta)
\begin{tikzpicture}[baseline=-0.65ex,scale=0.6]
    \draw (-0.6,-0.6) -- (-0.6,0.6) -- (0.6,0.6) -- (0.6,-0.6) -- (-0.6,-0.6);
    \node at (-1,0) (a1) {$a$};
       \node at (0,-1) (a2) {$e$};
           \node at (1,0) (a3) {$c$};
    \node at (0,1) (a4) {$d$};
\end{tikzpicture}(-\theta)
=\delta_{bd}\ .
\label{bui}
}
Putting \eqref{kha} and \eqref{bui} together we find the kink version of the QFT unitarity condition \eqref{qft}
\EQ{
\sum_e
\begin{tikzpicture}[baseline=-0.65ex,scale=0.6]
    \draw (-0.6,-0.6) -- (-0.6,0.6) -- (0.6,0.6) -- (0.6,-0.6) -- (-0.6,-0.6);
    \node at (-1,0) (a1) {$a$};
       \node at (0,-1) (a2) {$b$};
           \node at (1,0) (a3) {$c$};
    \node at (0,1) (a4) {$e$};
\end{tikzpicture}(\theta)
\begin{tikzpicture}[baseline=-0.65ex,scale=0.6]
    \draw (-0.6,-0.6) -- (-0.6,0.6) -- (0.6,0.6) -- (0.6,-0.6) -- (-0.6,-0.6);
    \node at (-1,0) (a1) {$a$};
       \node at (0,-1) (a2) {$d$};
           \node at (1,0) (a3) {$c$};
    \node at (0,1) (a4) {$e$};
\end{tikzpicture}(\theta)^*
=\delta_{bd}\ ,\qquad (\theta\text{ real})\ .
\label{unit}
}

\noindent{\bf Yang-Baxter Equation}\nopagebreak

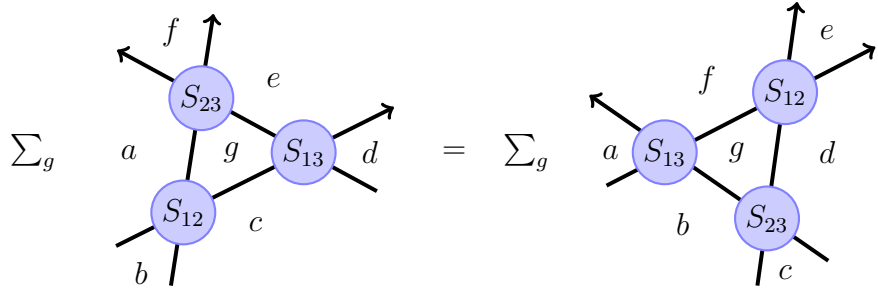
\begin{figure}
\begin{center}
\begin{tikzpicture} [scale=0.8,line width=1.5pt,inner sep=2mm,
place/.style={circle,draw=blue!50,fill=blue!20,thick}]
\begin{pgfonlayer}{top layer}
\node at (1.5,1.5) (pm1) {$S_{12}$}; 
\node at (1.8,3.4) (pm2) {$S_{23}$}; 
\node at (3.5,2.5) (m3) {$S_{13}$}; 
\node at (9.5,2.5) (m4) {$S_{13}$}; 
\node at (11.2,1.4) (m5) {$S_{23}$}; 
\node at (11.5,3.5) (m6) {$S_{12}$}; 
\end{pgfonlayer}
\begin{pgfonlayer}{foreground layer}
\node at (1.5,1.5) [place] (sm1) {\phantom{i}}; 
\node at (1.8,3.4) [place] (sm2) {\phantom{i}}; 
\node at (3.5,2.5) [place] (sm3) {\phantom{i}}; 
\node at (9.5,2.5) [place] (sm4) {\phantom{i}}; 
\node at (11.2,1.4) [place] (sm5) {\phantom{i}}; 
\node at (11.5,3.5) [place] (sm6) {\phantom{i}}; 
\end{pgfonlayer}
\node at (6,2.5) {$=$};
\node at (11,0) (i1) {};
\draw[->] (i1) -- (11.7,5);
\node at (12.5,0.5) (i2) {};
\draw[->] (i2) -- (8.25,3.45);
\node at (8.25,1.8) (i3) {};
\draw[->] (i3) -- (13,4.25) ;
\node at (5,1.7) (i4) {};
\draw[->] (i4) -- (0.4,4.2);
\node at (0.1,0.8) (i5) {};
\draw[->] (i5) -- (5,3.25);
\node at (1.25,0) (i6) {};
\draw[->] (i6) -- (2,4.8);
\node at (-1,2.5) {$\sum_g$};
\node at (7.2,2.5) {$\sum_g$};
\node at (0.6,2.5) {$a$};
\node at (2.3,2.5) {$g$};
\node at (4.6,2.5) {$d$};
\node at (8.6,2.5) {$a$};
\node at (10.7,2.5) {$g$};
\node at (12.2,2.5) {$d$};
\node at (0.8,0.5) {$b$};
\node at (11.5,0.5) {$c$};
\node at (2.7,1.3) {$c$};
\node at (9.8,1.3) {$b$};
\node at (1.3,4.5) {$f$};
\node at (12.2,4.5) {$e$};
\node at (3.0,3.7) {$e$};
\node at (10.2,3.7) {$f$};
\end{tikzpicture}
\caption{\small 
The Yang-Baxter equation in the IRF picture. $a,b,c,d,e,f$ label fixed vacua, while the internal vacuum $g$ should be summed over.}
\label{figybe}
\end{center}
\end{figure}

Finally, we briefly describe the Yang-Baxter equation in the IRF picture. As in the vertex picture it is easiest to represent it graphically, shown in Figure \ref{figybe}. The labels $a,b,c,d,e,f$ denote fixed vacua, while the internal vacuum $g$ should be summed over (in direct analogy with the sum of the indices labelling the internal lines in the usual Yang-Baxter equation). It can be checked that the kink S-matrix satisfies this version of the Yang-Baxter equation.\footnote{An alternative diagrammatical representation is given by the hexagon picture \cite{Baxter}, gotten from Figure \ref{figybe} by considering the dual graphs so that the vacua now label vertices.}

In principle one could check the Yang-Baxter equation in the IRF picture for a large class of external vacua. However, the kink S-matrix \eqref{web} can be thought of as a $4 \times 4$ matrix, depending not only on the difference of rapidities, but also say on the right vacuum (labelled as $c$ in \eqref{gtw}). Relabelling the external vacuum $d$ as $j$, Figure \ref{figybe} can then be understood as a deformed version of the usual Yang-Baxter equation (see Figure \ref{f2a}), known as the dynamical Yang-Baxter equation \cite{Felder:1994be}
\EQ{
S_{12}(\theta_1,\theta_2,j+\tfrac12\hat h_3) &S_{13}(\theta_1,\theta_3,j)S_{23}(\theta_2,\theta_3,j+\tfrac12\hat h_1)=
\\& S_{23}(\theta_2,\theta_3,j)S_{13}(\theta_1,\theta_3,j+\tfrac12\hat h_2)S_{12}(\theta_1,\theta_2,j)\ .
\label{dybe}
}
The operator $\hat h_i$ acts on the kink with rapidity $\theta_i$ as follows:
\EQ{
\hat h_i \, \ket{K_{ab}(\theta_i)} = 2(a-b)\ket{K_{ab}(\theta_i)}\ .
}
Note that $\hat h$ always acts on the kink that is not involved in the relevant scattering process---in practice it accounts for the fact that the same two-particle S-matrix can have different right vacua in different terms of the Yang-Baxter equation.

It is of particular interest that the kink S-matrix satisfies a dynamical Yang-Baxter equation as the latter's semi-classical expansion leads to a deformed version of the standard classical Yang-Baxter equation \cite{Felder:1994be}. This appears to partially explain the apparent anomaly noticed in the perturbative computation of various symmetric space sine-Gordon model S-matrices \cite{Hoare:2010fb,Hoare:2011fj}, discussed in detail in section \ref{slim}.

\section{The \texpdf{$\boldsymbol q$}{q}-Deformed World-Sheet S-matrix}\label{s3}

The S-matrix is constructed using a product of two copies of the fundamental $R$-matrix of the quantum group deformation of the triply extended superalgebra $\mh=\mpsu(2|2)\ltimes{\mathbb R}^3$ in \cite{Beisert:2008tw}, with the central extensions identified. Concentrating on a single $R$-matrix factor each particle multiplet is $4$ dimensional with two bosonic and two fermionic states, denoted here as $\{|\phi_m\rangle,|\psi_m\rangle\}$, where $m=\pm\frac12$ are the $j_z$ quantum  numbers for the two $\msu(2)$ bosonic subalgebras of $\mpsu(2|2)$. In general the theory is non-relativistic and therefore the kinematics is rather exotic as described in detail in Appendix \ref{A2}. States can be labelled by their rapidity which determines the velocity as usual by $v=\tanh\theta$. However, there is a maximum rapidity and for each $\theta$ there are two distinct physical states with different energy and momentum. In the following we will leave the choice of rapidity branch implicit.

The two-body S-matrix has non-vanishing elements
\EQ{
& \ket{\phi_m\phi_m} \longrightarrow A\ket{\phi_m\phi_m}\ ,\qquad   \ket{\psi_m\psi_m} \longrightarrow D\ket{\psi_m\psi_m}\ ,\\
&  \ket{\phi_{\pm\frac12}\phi_{\mp\frac12}} \longrightarrow \frac1{[2]}\Big((A-B)\ket{\phi_{\mp\frac12}\phi_{\pm\frac12}}+
(q^{\pm1} A+q^{\mp1}B)\ket{\phi_{\pm\frac12}\phi_{\mp\frac12}}\\ &\qquad\qquad\qquad\qquad\qquad\qquad\qquad\quad\;\;\;
+q^{\mp1}C\ket{\psi_{\pm\frac12}\psi_{\mp\frac12}}-C\ket{\psi_{\mp\frac12}\psi_{\pm\frac12}}\Big)\ ,\\
&  \ket{\psi_{\pm\frac12}\psi_{\mp\frac12}} \longrightarrow \frac1{[2]}\Big((D-E)\ket{\psi_{\mp\frac12}\psi_{\pm\frac12}}
+(q^{\pm1}D+q^{\mp1}E)\ket{\psi_{\pm\frac12}\psi_{\mp\frac12}}\\ &\qquad\qquad\qquad\qquad\qquad\qquad\qquad\qquad\;
+q^{\mp1}C\ket{\phi_{\pm\frac12}\phi_{\mp\frac12}}-C\ket{\phi_{\mp\frac12}\phi_{\pm\frac12}}\Big)\ ,\\
&  \ket{\phi_m\psi_n} \longrightarrow G\ket{\psi_n\phi_m}+H\ket{\phi_m\psi_n}\ ,\qquad
  \ket{\psi_m\phi_n} \longrightarrow H\ket{\psi_m\phi_n}+L\ket{\phi_n\psi_m}\ .
\label{jjs}
}
The functions $A=A(\theta_1,\theta_2)$, etc., are defined in \cite{Hoare:2011wr} (taken from the original reference \cite{Beisert:2008tw}). As the theory is generally not relativistically invariant the S-matrix is not a function of the difference $\theta_1-\theta_2$. Below we write the functions, including the dressing phase $v(\theta_1,\theta_2)$, in terms of the quantities $x^\pm_i=x^\pm(\theta_i)$ defined in Appendix \ref{A2}
\EQ{
A&=v\frac{U_1V_1}{U_2V_2}\cdot\frac{x_2^+-x_1^-}{x_2^--x_1^+}\ ,\qquad \qquad \,
B=A\Big(1-(1+q^{-2})\cdot\frac{x_2^+-x_1^+}{x_2^+-x_1^-}\cdot\frac{x_2^--\frac1{x_1^+}}{x_2^--\frac1{x_1^-}}\Big)\ ,\\
C&=iv(1+q^{-2})\Big(\frac{U_1V_1}{U_2V_2}\Big)^{3/2}\cdot\frac{1-\frac{x_2^+}{x_1^+}}{x_2^--\frac1{x_1^-}}\cdot\frac{\sqrt{(x_1^+-x_1^-)(x_2^+-x_2^-)}}{x_2^--x_1^+}\  ,\\
D&=-v\ ,\qquad \qquad \qquad \qquad\qquad \,
 E=D\Big(1-\frac{1+q^{-2}}{U_2^2V_2^2}\cdot\frac{x_2^+-x_1^+}{x_2^--x_1^+}\cdot
\frac{x_2^+-\frac1{x_1^-}}{x_2^--\frac1{x_1^-}}\Big)\ , \\
G&=vq^{-1/2}\frac1{U_2V_2}\cdot\frac{x_2^+-x_1^+}{x_2^--x_1^+}\ ,\qquad
L=vq^{1/2}U_1V_1\cdot\frac{x_2^--x_1^-}{x_2^--x_1^+}\ ,\\
H&=v\sqrt{\frac{U_1V_1}{U_2V_2}}\cdot\frac{\sqrt{(x_1^+-x_1^-)(x_2^+-x_2^-)}}{x_2^--x_1^+}\ .
\label{eqnsb}
}
The dressing phase was constructed in \cite{Hoare:2011wr}. Here we are implicitly considering the $q$-deformation of the usual magnon dressing phase (denoted $\sigma$ in \cite{Hoare:2011wr}). The single-particle quantities $U(\theta)$ and $V(\theta)$ are defined by
\EQ{
U^2=q^{-1}\frac{x^++\xi}{x^-+\xi}=q\frac{\frac1{x^-}+\xi}{\frac1{x^+}+\xi}\ ,\qquad
V^2=q^{-1}\frac{\xi x^++1}{\xi x^-+1}=q\frac{\frac\xi{x^-}+1}{\frac\xi{x^+}+1}\ ,
\label{jww}
}
where
\EQ{
q=\exp(i\pi/k)\ ,\qquad\xi=\frac{2g\sin(\pi/k)}{\sqrt{1+4g^2\sin^2(\pi/k)}}\ .
}

Physical states with real rapidity satisfy the reality condition $(x^\pm)^*=x^\mp$ which corresponds to real rapidity $\theta$. If the rapidity is continued into the complex plane the functions $A,B$, etc., satisfies a reality condition of the form
\EQ{
A(\theta_1,\theta_2)^*=A(\theta_2^*,\theta_1^*)\ .
}
Using these reality conditions, one can verify that the S-matrix is almost Hermitian analytic but, as in the $\msu(2)$ case described in section \ref{s2}, this is spoiled by the reflection amplitudes involving four bosons or four fermions:
\EQ{
\begin{tikzpicture}[baseline=-0.65ex,scale=0.5]
\filldraw[black] (0,0) circle (4pt);
    \draw[->] (-0.6,-0.6) -- (0.6,0.6);
    \draw[<-] (-0.6,0.6) -- (0.6,-0.6);
    \node at (-1.3,1.3) (a1) {$\phi_{\pm\frac12}$};
       \node at (1.3,-1.3) (a2) {$\phi_{\mp\frac12}$};
           \node at (1.3,1.3) (a3) {$\phi_{\mp\frac12}$};
    \node at (-1.3,-1.3) (a4) {$\phi_{\pm\frac12}$};
\end{tikzpicture}=\frac1{[2]}\big(q^{\pm1} A+q^{\mp1}B\big)\ ,\qquad
\begin{tikzpicture}[baseline=-0.65ex,scale=0.5]
\filldraw[black] (0,0) circle (4pt);
    \draw[->] (-0.6,-0.6) -- (0.6,0.6);
    \draw[<-] (-0.6,0.6) -- (0.6,-0.6);
    \node at (-1.3,1.3) (a1) {$\psi_{\pm\frac12}$};
       \node at (1.3,-1.3) (a2) {$\psi_{\mp\frac12}$};
           \node at (1.3,1.3) (a3) {$\psi_{\mp\frac12}$};
    \node at (-1.3,-1.3) (a4) {$\psi_{\pm\frac12}$};
\end{tikzpicture}=\frac1{[2]}\big(q^{\pm1}D+q^{\mp1}E\big)\ .
}
The lessons from the $\msu(2)$ quantum group example in section \ref{s2} suggests that Hermitian analyticity should be manifest in a kink basis obtained by making the vertex-to-IRF transformation on the bosonic $\msu(2)\oplus\msu(2)$ subalgebra of $\mpsu(2|2)$. In the $\msu(2)$ example a key identity which ensures the overall consistency of the transformation is \eqref{kid} and one can verify that this also holds in the present case, in both the bosonic and fermionic sector. In each sector, involving four bosonic or four fermionic particles, we have the same structure of the S-matrix as in \eqref{wea}--\eqref{wea2} with
\EQ{
&S_I=A\ ,\qquad S_T=\frac1{[2]}(A-B)\ ,\qquad S_R^\pm=\frac1{[2]}(q^{\pm1}A+q^{\mp1}B)\ ,
\\
&\widetilde S_I=D\ ,\qquad \widetilde S_T=\frac1{[2]}(D-E)\ ,\qquad \widetilde S_R^\pm=\frac1{[2]}(q^{\pm1}D+q^{\mp1}E)\ ,
}
In both case the crucial identity \eqref{kid} holds. Of course it had to be so because of the $\msu(2)\oplus\msu(2)$ quantum group invariance of the S-matrix.

The above identity is important because it means that we can use the same formulae for the S-matrix used in the $\msu(2)$ case given in Appendix \ref{A1} with the appropriate values of $S_I$, $S_T$ and $S_R^\pm$. In the kink picture, the vacua are labelled by a pair of spins $(j,l)$, one for each $\msu(2)$, and kinks of the form $K_{j,j\pm\frac12}^{l,l}$ are bosonic while $K_{j,j}^{l,l\pm\frac12}$ are fermionic.\footnote{For the world-sheet S-matrix, states come in a tensor product of two copies of the $\mpsu(2|2)$ $R$-matrix. Therefore the vacua are labelled by four spins associated to the bosonic $\msu(2)^{\oplus4}$.} Making the vertex-to-IRF transformation as in section \ref{s2} gives the kink S-matrix
\EQ{
 \ket{K^{l,l}_{j\pm 1,j\pm\frac12}K^{l,l}_{j\pm\frac12,j}} \longrightarrow & \ A\ket{K^{l,l}_{j\pm 1,j\pm\frac12}K^{l,l}_{j\pm\frac12,j}}\ ,\\
 \ket{K_{j,j}^{l\pm1,l\pm\frac12}K_{j,j}^{l\pm\frac12,l}} \longrightarrow & \ D\ket{K_{j,j}^{l\pm1,l\pm\frac12}K_{j,j}^{l\pm\frac12,l}}\ ,\\
 \ket{K^{l,l}_{j,j\pm\frac12}K^{l,l}_{j\pm\frac12,j}} \longrightarrow & \ \frac{[2j+1\mp1]A+[2j+1\pm1]B}{[2][2j+1]}\ket{K^{l,l}_{j,j\pm\frac12}K^{l,l}_{j\pm\frac12,j}}\\ &+
\frac{\sqrt{[2j+2][2j]}}{[2][2j+1]}(A-B)\ket{K^{l,l}_{j,j\mp\frac12}K^{l,l}_{j\mp\frac12,j}}\\
&+\sqrt{\frac{[2j+1\pm1][2l+1\pm1]}{[2j+1][2l+1]}}\frac{C}{[2]}\ket{K_{j,j}^{l,l\pm\frac12}K_{j,j}^{l\pm\frac12,l}}\\ &-\sqrt{\frac{[2j+1\pm1][2l+1\mp1]}{[2j+1][2l+1]}}\frac{C}{[2]}\ket{K_{j,j}^{l,l\mp\frac12}K_{j,j}^{l\mp\frac12,l}}\ ,\\
 \ket{K_{j,j}^{l,l\pm\frac12}K_{j,j}^{l\pm\frac12,l}} \longrightarrow & \ \frac{[2l+1\mp1]D+[2l+1\pm1]E}{[2][2l+1]}\ket{K_{j,j}^{l,l\pm\frac12}K_{j,j}^{l\pm\frac12,l}}\\ &+
\frac{\sqrt{[2l+2][2l]}}{[2][2l+1]}(D-E)\ket{K_{j,j}^{l,l\mp\frac12}K_{j,j}^{l\mp\frac12,l}}
\ ,\\
&+\sqrt{\frac{[2j+1\pm1][2l+1\pm1]}{[2j+1][2l+1]}}\frac{C}{[2]}\ket{K_{j,j\pm\frac12}^{l,l}K_{j\pm\frac12,j}^{l,l}}\\ &-\sqrt{\frac{[2j+1\mp1][2l+1\pm1]}{[2j+1][2l+1]}}\frac{C}{[2]}\ket{K_{j,j\mp\frac12}^{l,l}K_{j\mp\frac12,j}^{l,l}}\ ,\\
 \ket{K_{j\pm\frac12,j}^{l\mp\frac12,l\mp\frac12}K_{j,j}^{l\mp\frac12,l}} \longrightarrow & \
G\ket{K_{j\pm\frac12,j\pm\frac12}^{l\mp\frac12,l}K_{j\pm\frac12,j}^{l,l}}
+H\ket{K_{j\pm\frac12,j}^{l\mp\frac12,l\mp\frac12}K_{j,j}^{l\mp\frac12,l}}\ ,\\
 \ket{K_{j\mp\frac12,j\mp\frac12}^{l\pm\frac12,l}K_{j\mp\frac12,j}^{l,l}} \longrightarrow & \
H\ket{K_{j\mp\frac12,j\mp\frac12}^{l\pm\frac12,l}K_{j\mp\frac12,j}^{l,l}}
+L\ket{K_{j\mp\frac12,j}^{l\pm\frac12,l\pm\frac12}K_{j,j}^{l\pm\frac12,l}}\ ,
\label{web2}
}

These S-matrix elements can be summarized as follows. Firstly kinks $K_{ab}^{uv}(\theta)$ must have either $|a-b|=\frac12$ and $u=v$, or $a=b$ and $|u-v|=\frac12$---the former being a boson and the latter a fermion. We introduce the notation
\EQ{
\begin{tikzpicture}[baseline=-0.65ex,scale=0.6]
    \draw (-0.6,-0.6) -- (-0.6,0.6) -- (0.6,0.6) -- (0.6,-0.6) -- (-0.6,-0.6);
    \node at (-1.3,0) (a1) {$a,u$};
              \node at (0,-1) (a2) {$b,v$};
           \node at (1.3,0) (a3) {$c,w$};
    \node at (0,1) (a4) {$d,y$};
\end{tikzpicture}(\theta_1,\theta_2)\ ,
\label{dek}
}
for the process $K_{ab}^{uv}(\theta_1)+K_{bc}^{vw}(\theta_2)\longrightarrow
K_{ad}^{uy}(\theta_1)+K_{dc}^{yw}(\theta_2)$.
The processes involving $BB\to BB$ can then be written in the compact form
\EQ{
\begin{tikzpicture}[baseline=-0.65ex,scale=0.6]
   \draw (-0.6,-0.6) -- (-0.6,0.6) -- (0.6,0.6) -- (0.6,-0.6) -- (-0.6,-0.6);
    \node at (-1.3,0) (a1) {$a,u$};
              \node at (0,-1) (a2) {$b,u$};
           \node at (1.3,0) (a3) {$c,u$};
    \node at (0,1) (a4) {$d,u$};
\end{tikzpicture}=S_I\delta_{bd}+e^{i\pi(a+c-b-d)}S_T\sqrt{\frac{[2b+1][2d+1]}{[2a+1][2c+1]}}\delta_{ac}\ ,
\label{gtw2}
}
while those involving $FF\to FF$ are given by
\EQ{
\begin{tikzpicture}[baseline=-0.65ex,scale=0.6]
   \draw (-0.6,-0.6) -- (-0.6,0.6) -- (0.6,0.6) -- (0.6,-0.6) -- (-0.6,-0.6);
    \node at (-1.3,0) (a1) {$a,u$};
              \node at (0,-1) (a2) {$a,v$};
           \node at (1.3,0) (a3) {$a,w$};
    \node at (0,1) (a4) {$a,y$};
\end{tikzpicture}=\widetilde S_I\delta_{vy}+e^{i\pi(u+w-y-v)}\widetilde S_T\sqrt{\frac{[2v+1][2y+1]}{[2u+1][2w+1]}}\delta_{uw}\ .
\label{gtw3}
}
For those involving $BB\to FF$ or $FF\to BB$ we have
\EQ{
\begin{tikzpicture}[baseline=-0.65ex,scale=0.6]
   \draw (-0.6,-0.6) -- (-0.6,0.6) -- (0.6,0.6) -- (0.6,-0.6) -- (-0.6,-0.6);
    \node at (-1.3,0) (a1) {$a,u$};
              \node at (0,-1) (a2) {$b,u$};
           \node at (1.3,0) (a3) {$a,u$};
    \node at (0,1) (a4) {$a,v$};
\end{tikzpicture}=\begin{tikzpicture}[baseline=-0.65ex,scale=0.6]
   \draw (-0.6,-0.6) -- (-0.6,0.6) -- (0.6,0.6) -- (0.6,-0.6) -- (-0.6,-0.6);
    \node at (-1.3,0) (a1) {$a,u$};
              \node at (0,-1) (a2) {$a,v$};
           \node at (1.3,0) (a3) {$a,u$};
    \node at (0,1) (a4) {$b,u$};
\end{tikzpicture}=
-e^{i\pi(a-b+u-v)}\frac C{[2]}\sqrt{\frac{[2b+1][2v+1]}{[2a+1][2u+1]}}\ .
\label{gtw4}
}
Finally there are the processes involving $BF\to FB+BF$ and $FB\to BF+FB$, respectively,
\EQ{
\begin{tikzpicture}[baseline=-0.65ex,scale=0.6]
   \draw (-0.6,-0.6) -- (-0.6,0.6) -- (0.6,0.6) -- (0.6,-0.6) -- (-0.6,-0.6);
    \node at (-1.3,0) (a1) {$a,u$};
              \node at (0,-1) (a2) {$b,u$};
           \node at (1.3,0) (a3) {$b,v$};
    \node at (0,1) (a4) {$d,y$};
\end{tikzpicture}=G\delta_{ad}\delta_{vy}+H\delta_{bd}\delta_{uy}\ ,\qquad
\begin{tikzpicture}[baseline=-0.65ex,scale=0.6]
   \draw (-0.6,-0.6) -- (-0.6,0.6) -- (0.6,0.6) -- (0.6,-0.6) -- (-0.6,-0.6);
    \node at (-1.3,0) (a1) {$a,u$};
              \node at (0,-1) (a2) {$a,v$};
           \node at (1.3,0) (a3) {$b,v$};
    \node at (0,1) (a4) {$d,y$};
\end{tikzpicture}=H\delta_{ad}\delta_{vy}+L\delta_{bd}\delta_{uy}\ .
\label{gtw5}
}

The functions involved in the S-matrix satisfy the crossing symmetry relations
\EQ{
&S_I(\theta_1,\theta_2)=S_T(i\pi+\theta_2,\theta_1)\ ,\qquad
\widetilde S_I(\theta_1,\theta_2)=\widetilde S_T(i\pi+\theta_2,\theta_1)\ ,\\ &
C(\theta_1,\theta_2)=[2]H(i\pi+\theta_2,\theta_1)\ ,\qquad G(\theta_1,\theta_2)=
L(i\pi+\theta_2,\theta_1)\ .
}
Using these identities,
the kink S-matrix in \eqref{dek} is seen to be crossing symmetric if 
we define charge conjugation as\footnote{Recall that in this superalgebra case we should use the supertranspose as oppose to the transpose. This contributes a factor of $e^{2i\pi(v+y-u-w)}$ to the crossing equation.}
\EQ{
{\cal C}\ket{K_{ab}^{uv}(\theta)}= e^{i\pi(b-a+v-u)}\sqrt{\frac{[2a+1][2u+1]}{[2b+1][2v+1]}} \ket{K_{ba}^{vu}(\theta)} \ .
}

One can check that the S-matrix satisfies Hermitian analyticity in the kink picture
\EQ{
\begin{tikzpicture}[baseline=-0.65ex,scale=0.6]
    \draw (-0.6,-0.6) -- (-0.6,0.6) -- (0.6,0.6) -- (0.6,-0.6) -- (-0.6,-0.6);
    \node at (-1.3,0) (a1) {$a,u$};
              \node at (0,-1) (a2) {$b,v$};
           \node at (1.3,0) (a3) {$c,w$};
    \node at (0,1) (a4) {$d,y$};
\end{tikzpicture}(\theta_1^*,\theta_2^*)^*
=
\begin{tikzpicture}[baseline=-0.65ex,scale=0.6]
    \draw (-0.6,-0.6) -- (-0.6,0.6) -- (0.6,0.6) -- (0.6,-0.6) -- (-0.6,-0.6);
    \node at (-1.3,0) (a1) {$a,u$};
              \node at (0,-1) (a2) {$d,y$};
           \node at (1.3,0) (a3) {$c,w$};
    \node at (0,1) (a4) {$b,v$};
\end{tikzpicture}(\theta_2,\theta_1)\ .
\label{bui1}}
In the kink basis, the braiding unitarity relation takes the form
\EQ{
\sum_{e,z}
\begin{tikzpicture}[baseline=-0.65ex,scale=0.6]
    \draw (-0.6,-0.6) -- (-0.6,0.6) -- (0.6,0.6) -- (0.6,-0.6) -- (-0.6,-0.6);
    \node at (-1.3,0) (a1) {$a,u$};
              \node at (0,-1) (a2) {$b,v$};
           \node at (1.3,0) (a3) {$c,w$};
    \node at (0,1) (a4) {$e,z$};
\end{tikzpicture}(\theta_1,\theta_2)
\begin{tikzpicture}[baseline=-0.65ex,scale=0.6]
    \draw (-0.6,-0.6) -- (-0.6,0.6) -- (0.6,0.6) -- (0.6,-0.6) -- (-0.6,-0.6);
    \node at (-1.3,0) (a1) {$a,u$};
              \node at (0,-1) (a2) {$e,z$};
           \node at (1.3,0) (a3) {$c,w$};
    \node at (0,1) (a4) {$d,y$};
\end{tikzpicture}(\theta_2,\theta_1)
=\delta_{bd}\delta_{vy}\ .
\label{bui2}
}
Putting \eqref{bui1} and \eqref{bui2} together then gives the kink version of the QFT unitarity condition \eqref{unit}
\EQ{
\sum_{e,z}
\begin{tikzpicture}[baseline=-0.65ex,scale=0.6]
    \draw (-0.6,-0.6) -- (-0.6,0.6) -- (0.6,0.6) -- (0.6,-0.6) -- (-0.6,-0.6);
    \node at (-1.3,0) (a1) {$a,u$};
              \node at (0,-1) (a2) {$b,v$};
           \node at (1.3,0) (a3) {$c,w$};
    \node at (0,1) (a4) {$e,z$};
\end{tikzpicture}(\theta_1,\theta_2)
\begin{tikzpicture}[baseline=-0.65ex,scale=0.6]
    \draw (-0.6,-0.6) -- (-0.6,0.6) -- (0.6,0.6) -- (0.6,-0.6) -- (-0.6,-0.6);
    \node at (-1.3,0) (a1) {$a,u$};
              \node at (0,-1) (a2) {$d,y$};
           \node at (1.3,0) (a3) {$c,w$};
    \node at (0,1) (a4) {$e,z$};
\end{tikzpicture}(\theta_1,\theta_2)^*
=\delta_{bd}\delta_{vy}\ ,\qquad(\theta_i \text{ real})\ .
\label{bui3}
}

Finally, it can be checked that \eqref{web2} satisfies both the Yang-Baxter equation in the IRF picture and also the dynamical Yang-Baxter equation given by the obvious generalization of the $\msu(2)$ case---see Figure \ref{figybe} and equation \eqref{dybe} respectively.

\section{Limits of the Kink S-Matrix\label{slim}}

In this section we discuss two limits of the kink S-matrix. In the first we expect to recover the string S-matrix, while the second is conjectured to be related to the Pohlmeyer-reduced $\text{AdS}_5 \times S^5$ superstring.

{
Taking the string limit ($k \rightarrow \infty$) of the vertex S-matrix \eqref{jjs} we recover the S-matrix of the light-cone gauge-fixed string theory \cite{Beisert:2008tw}---in this limit the vertex S-matrix is QFT unitary \cite{Arutyunov:2009ga}. However, taking this limit in the kink S-matrix we find additional $(j,l)$-dependent factors originating from the $q$-numbers in equation \eqref{web2}, and to recover the string S-matrix, we also need to take the $\msu(2)$ spins $j$ and $l$ to infinity. 

The limit $j,l\to\infty$ can be justified as follows.
Recall that, in the kink picture, $(j,l)$ label the vacua so that, for $q=e^{i\pi/k}$,
\EQ{
j,l\in\Big\{0,\frac12,1,\ldots, \frac{k}2-1\Big\}
}
and, in the $k\to\infty$ limit, the kinks interpolate between an infinite number of vacua. Naively, therefore, this limit distinguishes between the vacua $j,l=0$ and $j,l=\frac{k}2-1\to\infty$, but it is worth noticing that the S-matrix elements~\eqref{gtw} and~\eqref{gtw2}--\eqref{gtw5} are invariant if we change all the labels corresponding to the vacua according to
$j\to \tfrac{k}2-1-j$ and $l\to \tfrac{k}2-1-l$. This suggests to take the $k\to\infty$ limit keeping the symmetry between $j,l=0$ and $j,l=\frac{k}2-1\to\infty$. In order to do that, we redefine the spins
\begin{equation}\label{rdef}
j = \frac{k-2}4 + \tilde \jmath \ , \qquad
l = \frac{k-2}4 + \tilde l \ ,
\end{equation}
and then take the $k\to\infty$ limit keeping $\tilde \jmath$ and $\tilde l$ fixed. This implies $j,l\to\infty$, and the dependence on $\tilde \jmath$ 
and $\tilde l$ drops out leaving the S-matrix of the light-cone gauge-fixed string theory. Notice that this limit gives rise to an infinite number of vacua labelled by $\tilde \jmath, \tilde l$ which play the role of topological charges.
}

The relativistic limit ($g \rightarrow \infty$) in the vertex picture is known to give a relativistic S-matrix related to the Pohlmeyer reduction of the $\text{AdS}_5 \times S^5$ superstring \cite{Hoare:2011fj,Hoare:2011nd}. However, the large $k$ expansion of the vertex S-matrix does not agree with the perturbative computation of \cite{Hoare:2009fs,Hoare:2011fj}. This might be expected as the perturbative S-matrix does not satisfy the Yang-Baxter equation. In particular, the tree-level result does not satisfy the classical Yang-Baxter equation, which follows from the quantum one \eqref{ybe} assuming the S-matrix has the form ${\cal P} + \tfrac 1k \, {\cal T}$, where ${\cal P}$ is the permutation operator. Furthermore, the perturbative S-matrix is unitary while the vertex S-matrix is not.

In sections \ref{s2} and \ref{s3}, it was shown that starting from the vertex S-matrix one can use the vertex-to-IRF transformation to move to the kink picture, in which both QFT unitarity and the Yang-Baxter equation are satisfied. Therefore, it is natural to ask whether the large $k$ expansion of the relativistic limit of the kink S-matrix \eqref{web2} has any relation to the perturbative S-matrix of the Pohlmeyer-reduced theory. For reference we give the relativistic limit of the functions parametrizing the S-matrix:\footnote{
To facilitate comparison with the perturbative computation \cite{Hoare:2011fj} we have extracted a factor from the phase
\EQ{\tilde v = v \, \sinh \frac\theta 2 \csch\left(\frac\theta 2 + \frac{i\pi}{2k}\right)  \ .}
The dressing phase in the relativistic limit is given in \cite{Hoare:2011fj} and in integral form in \cite{Hoare:2011nd}.}
\EQ{
& A(\theta) = \tilde v \, \csch\frac\theta2 \sinh\left(\frac\theta 2 - \frac{i\pi}{2k}\right) \ ,\quad \ \ \,
D(\theta) = - \, \tilde v \, \csch\frac\theta2 \sinh\left(\frac\theta 2 + \frac{i\pi}{2k}\right) \ , \\
& B(\theta) = - 2 \, i \, \tilde v \, \csch\theta \left(\sin\frac{\pi }{2 k}-i \cosh\left(\frac{\theta }{2}+\frac{3 i \pi }{2 k}\right)\sinh\frac{\theta }{2}\right) \ , \\
& E(\theta) = - 2 \, i \, \tilde v \, \csch\theta \left(\sin\frac{\pi }{2 k}+i \cosh\left(\frac{\theta }{2}-\frac{3 i \pi }{2 k}\right)\sinh\frac{\theta }{2}\right) \ , \\
& C(\theta) = - \, 2 \,\tilde v \cos\frac{\pi }{k} \sin\frac{\pi }{2 k} \sech\frac{\theta }{2} \ , \quad \ \ \
H(\theta) = - \, i\, \tilde v\, \sin\frac{\pi }{2 k}\csch \frac \theta 2\ , \vphantom{\left(\frac\theta 2 \right)} \\
& G(\theta) = L(\theta) = \tilde v \ . \vphantom{\left(\frac\theta 2 \right)}
}
The S-matrix now depends only on the difference of rapidities, $\theta = \theta_1 - \theta_2$, as required by Lorentz symmetry. The relation between $x^\pm$ and the rapidity for arbitrary $(g,k)$ is discussed in Appendix \ref{A2}.

In addition to depending on $k$ and $g$ the kink S-matrix also depends on the $\msu(2)$ spins $j$ and $l$, which should be taken large for a good semi-classical interpretation (see Appendix \ref{A3}). However, the $(j,l) \rightarrow \infty$ limit is not well-defined for finite $k$.\footnote{As discussed in section \ref{s5}, for a consistent physical theory $k$ is required to be an integer while $j$ and $l$ should take values in the finite set $\{0,\tfrac12,\ldots,\tfrac k2-1\}$.} Our approach is therefore to first redefine the spins using \eqref{rdef} and then expand around large $k$. To the one-loop order we find the following S-matrix
\EQ{
\ket{K^{l,l}_{j\pm1,j\pm\frac12}K^{l,l}_{j\pm\frac12,j}} \longrightarrow
& \ \tilde v \, \big(1-\frac{i\pi}{2k}\coth\frac\theta2-\frac{\pi^2}{8k^2}\big)
\ket{K^{l,l}_{j\pm1,j\pm\frac12}K^{l,l}_{j\pm\frac12,j}} \ , \\
\ket{K_{j,j}^{l\pm1,l\pm\frac12}K_{j,j}^{l\pm\frac12,l}} \longrightarrow
& \ \tilde v \, \big(-1-\frac{i\pi}{2k}\coth\frac\theta2+\frac{\pi^2}{8k^2}\big)
\ket{K_{j,j}^{l\pm1,l\pm\frac12}K_{j,j}^{l\pm\frac12,l}} \ , \\
\ket{K^{l,l}_{j,j\pm\frac12}K^{l,l}_{j\pm\frac12,j}} \longrightarrow
& - \tilde v \, \big(\frac{i\pi}{k}\coth\theta \mp \frac{2\pi^2 \tilde \jmath}{k^2}-\frac{\pi^2}{2k^2}\big)
\ket{K^{l,l}_{j,j\pm\frac12}K^{l,l}_{j\pm\frac12,j}} \\
& + \tilde v \, \big(1+\frac{i\pi}{2k}\tanh\frac\theta 2-\frac{5\pi^2}{8k^2}\big)
\ket{K^{l,l}_{j,j\mp\frac12}K^{l,l}_{j\mp\frac12,j}} \\
& - \tilde v \, \big(\frac{\pi}{2k}\sech\theta\big)
\ket{K_{j,j}^{l,l\pm\frac12}K_{j,j}^{l\pm\frac12,l}} \\
& + \tilde v \, \big(\frac{\pi}{2k}\sech\theta\big)
\ket{K_{j,j}^{l,l\mp\frac12}K_{j,j}^{l\mp\frac12,l}} \ , \\
\ket{K_{j,j}^{l,l\pm\frac12}K_{j,j}^{l\pm\frac12,l}} \longrightarrow
& - \tilde v \, \big(\frac{i\pi}{k}\coth\theta \pm \frac{2\pi^2\tilde l}{k^2}+\frac{\pi^2}{2k^2}\big)
\ket{K_{j,j}^{l,l\pm\frac12}K_{j,j}^{l\pm\frac12,l}} \\
& - \tilde v \, \big(1-\frac{i\pi}{2k}\tanh\frac\theta2-\frac{5\pi^2}{8k^2}\big)
\ket{K_{j,j}^{l,l\mp\frac12}K_{j,j}^{l\mp\frac12,l}} \ , \\
& - \tilde v \, \big(\frac{\pi}{2k}\sech\theta\big)
\ket{K_{j,j\pm\frac12}^{l,l}K_{j\pm\frac12,j}^{l,l}} \\
& + \tilde v \, \big(\frac{\pi}{2k}\sech\theta\big)
\ket{K_{j,j\mp\frac12}^{l,l}K_{j\mp\frac12,j}^{l,l}} \ , \\
\ket{K_{j\pm\frac12,j}^{l\mp\frac12,l\mp\frac12}K_{j,j}^{l\mp\frac12,l}} \longrightarrow
& \ \tilde v \,
\ket{K_{j\pm\frac12,j\pm\frac12}^{l\mp\frac12,l}K_{j\pm\frac12,j}^{l,l}}
- \tilde v \, \big(\frac{i\pi}{2k}\csch\frac\theta 2\big)
\ket{K_{j\pm\frac12,j}^{l\mp\frac12,l\mp\frac12}K_{j,j}^{l\mp\frac12,l}} \ , \\
\ket{K_{j\mp\frac12,j\mp\frac12}^{l\pm\frac12,l}K_{j\mp\frac12,j}^{l,l}} \longrightarrow
& - \tilde v \, \big(\frac{i\pi}{2k}\csch\frac\theta 2\big)\ket{K_{j\mp\frac12,j\mp\frac12}^{l\pm\frac12,l}K_{j\mp\frac12,j}^{l,l}}
+ \tilde v \,
\ket{K_{j\mp\frac12,j}^{l\pm\frac12,l\pm\frac12}K_{j,j}^{l\pm\frac12,l}} \ .
\label{webex}}
Comparing with \cite{Hoare:2009fs,Hoare:2011fj} we see that at the tree level the expansion of the kink S-matrix matches the perturbative computation of the Pohlmeyer-reduced theory S-matrix.\footnote{Note that here we have redefined the fermions by a factor of $e^{i\pi/4}$ compared to \cite{Hoare:2009fs,Hoare:2011fj}.} Furthermore, setting $\tilde \jmath = \tilde l = 0$ in the one-loop terms we recover the real part of the perturbative computation. The exact agreement however breaks down at one-loop. This may be expected as it is only for large $k$ that the perturbative states are a good approximation for the kink states \cite{Hollowood:2011fq}.

While agreement with the perturbative S-matrix is no longer true at one-loop, the presence of the $\msu(2)$ spins in the one-loop amplitudes provides us with the explanation of the Yang-Baxter ``anomaly'' of the tree-level S-matrix \cite{Hoare:2009fs,Hoare:2010fb,Hoare:2011fj}. The identity that the tree-level S-matrix should satisfy is modified from the classical Yang-Baxter equation by a contribution originating from the shifts in vacua in the dynamical Yang-Baxter equation. This is the usual construction whereby one recovers the modified classical Yang-Baxter equation from the semi-classical expansion of the dynamical Yang-Baxter equation \cite{Felder:1994be}.

\noindent
\textbf{A simpler example}\nopagebreak

For completeness we also discuss the analogous construction for the $\msu(2)$ quantum group and the $S^5$ symmetric space sine-Gordon theory. In this case the Lie algebra underlying the symmetry of the theory is $\mso(4) \cong \msu(2) \oplus \msu(2)$ and the S-matrix should factorize accordingly into two copies of the $R$-matrix associated to the $\msu(2)$ quantum group of section \ref{s2}. The map between the parameters $x$ and $q$ and $\theta$ and $k$ is given in \eqref{de2}. It is therefore useful to define
\EQ{
\lambda = \frac{k+1}{k+2} \ , \qquad \omega = \frac{\pi}{k+2} \ ,}
so that
\EQ{
x = e^{\lambda \theta} \ , \qquad q = e^{i\omega}\ .}

The perturbative S-matrix for this theory was computed in \cite{Hoare:2010fb} to the one-loop order, or equivalently ${\cal O}(\omega^2)$. Pulling out an overall factor\footnote{This expression is motivated by various observations in \cite{Hollowood:2010rv} and Appendix G of \cite{Hoare:2011fj}. Furthermore it satisfies the braiding unitarity and crossing relations \eqref{phasecross} and $v(\theta) = v(i\pi-\theta)$. Note that the square root disappears in the tensor product.}
\EQ{
v(\theta) = & \frac{1}{2\sinh\big(\lambda \theta + i\omega\big)} 
\sqrt{\frac{\sinh\big(\frac{\lambda \theta}{2} + \frac{i\omega}{2}\big)\cosh\big(\frac{\lambda \theta}{2} - \frac{i\omega}{2}\big)}{\sinh\big(\frac{\lambda \theta}{2} - \frac{i\omega}{2}\big) \cosh\big(\frac{\lambda \theta}{2} + \frac{i\omega}{2}\big)}} \ \frac{\cosh\big(\frac{\lambda \theta}{2} + \frac{i\omega}{4}\big)}{\cosh\big(\frac{\lambda \theta}{2} - \frac{i\omega}{4}\big)}
\\ & \qquad \qquad \qquad \qquad \qquad \qquad \qquad \Big(1 - \frac{i\omega^2}{\pi} \lambda \theta \coth \lambda \theta \csch \lambda \theta + {\cal O}(k^{-3}) \Big)\ ,}
the perturbative result is given by
\EQ{
\mathbb S\ket{\phi_m(\theta_1) \phi_n(\theta_2)} = v(\theta)\Big[&\Big(2\sinh \lambda \theta + \frac{i\omega^2}{\pi} \big(1+(i \pi -\lambda  \theta \big) \sinh \lambda \theta\big) \Big) \delta_m^q \delta_n^p \\ & + \Big(2 i \omega  \cosh \lambda \theta +\frac{2 i \omega^2}{\pi } \lambda  \theta  \sinh\lambda \theta \Big)\delta_m^p \delta_n^q\Big]\ket{\phi_p(\theta_2)\phi_q(\theta_1)}\ .
\label{compare1}}

The expansion of the S-matrix associated to the $\msu(2)$ quantum group in the kink basis \eqref{web} to the same order is given by
\EQ{
&\ket{K_{j\pm1,j\pm\frac12}(\theta_1)K_{j\pm\frac12,j}(\theta_2)}\longrightarrow 
\\ & \hspace{15pt} v(\theta) \Big(2\sinh \lambda \theta +2 i \omega \cosh \lambda \theta - \omega^2 \sinh \lambda \theta \Big)\ket{K_{j\pm1,j\pm\frac12}(\theta_2)K_{j\pm\frac12,j}(\theta_1)}\ ,\\
&\ket{K_{j,j\pm\frac12}(\theta_1)K_{j\pm\frac12,j}(\theta_2)}\longrightarrow
\\ & \hspace{15pt} v(\theta) \Big(2\sinh \lambda \theta -\omega^2\sinh \lambda \theta \Big) \ket{K_{j,j\mp\frac12}(\theta_2)K_{j\mp\frac12,j}(\theta_1)}
\\ &\hspace{40pt}+v(\theta) \Big(2 i \omega  \cosh \lambda \theta \pm 4 \omega^2 \tilde \jmath \sinh\lambda \theta \Big) \ket{K_{j,j\pm\frac12}(\theta_2)K_{j\pm\frac12,j}(\theta_1)}\ .
\label{compare2}
}
Here we have first redefined the $\msu(2)$ spin 
\begin{equation}\label{rrdef}
j = \frac{k}4 + \tilde \jmath \ ,
\end{equation}
which is equivalent to \eqref{rdef} taking account of the shift in $k$, and then expanded around large $k$. Comparing \eqref{compare1} and \eqref{compare2} we again see that we recover the perturbative result at the tree level. Furthermore, if we set $\tilde \jmath = 0$, we find agreement with the real part of the perturbative computation at the one-loop level. However, in the imaginary part there is disagreement. Again, this may be expected as it is only for large $k$ that the perturbative states are a good approximation for the kink states \cite{Hollowood:2011fq}.

The disagreement stems from the presence of ${\cal O}(\tilde \jmath)$ terms in the one-loop reflection amplitudes. More precisely, they imply the crossing and unitarity relations that the one-loop amplitudes should satisfy are different to those that the perturbative result satisfies. Consequently, for the perturbative computation to match the expansion of the kink S-matrix beyond the tree level it would need to be modified to include the $\msu(2)$ spin $j$.

\section{The Restricted Theories}\label{s5}

When $q$ is generic it is known that representations of quantum groups are simple deformations of representations of the undeformed group.  {However, when $q$ is a root of unity, the case pertinent to our discussion, the representation theory of quantum groups is more subtle{, and we have summarized its main features for the case of $SU(2)$ in Appendix~\ref{a5}}. For $q=e^{i\pi/k}$, there are a set of irreducible ``good" representations of dimension $<k$, and the idea is to define a restricted theory by removing from the Hilbert space the remaining ``bad" representations. The ``good" representations are of Type A  and denoted $V_j^{(+1)}$ in Appendix \ref{a5}, with $j\leq k/2-1$. The ``bad" representations consists of a finite set of reducible but indecomposable representations of dimension $2k$, in addition to the irreducible representation $V_{j}^{(+1)}$ with $j=(k-1)/2$ of dimension~$k$.}\footnote{ {Notice that the q-CG coefficients~\eqref{qCG} blow up for $j=(k-1)/2$.}} 
It is then possible to define a restricted representation theory which only includes the ``good" representations and the  
way to do this in practice is to use the vertex-to-IRF change of basis to go to the kink basis and simply insist that the vacua lie in the finite set $\{0,\frac12,1,\ldots,\frac k2-1\}$. This consistently implements the restriction.

What is particularly nice about this restriction is that it meshes perfectly with the spectrum of bound states of the theory and the bootstrap procedure that determines the S-matrix elements of the bound states. With the vacua restricted to the set $\{0,\frac12,1,\ldots,\frac k2-1\}$ it is clear that states in the original vertex picture are restricted. For instance, the Hilbert space cannot contain the  states $\ket{\phi_m(\theta_1)\phi_m(\theta_2)\cdots\phi_m(\theta_N)}$ with $N>k-2$. The bound  {states} transform in the short representations $\langle a-1,0\rangle$ (the magnons) or $\langle0,a-1\rangle$ (the solitons) of $U_q(\mh)$.  {Taking the magnon bound states,
they contain states in the representation\footnote{The $\msu(2)$ labels here are twice the spin and the representation theory of $U_q(\mh)$ is discussed in \cite{Beisert:2008tw,Hoare:2011nd}.} $(a,0)\oplus(a-1,1)\oplus(a-2,0)$ of the subalgebra $U_q(\msu(2))\times U_q(\msu(2))$ and, clearly, only states with $a\leq k$ can appear in the spectrum of the restricted model.} Note also that the bound states near the top of the tower, those with $a=k$ and $k-1$, have a modified  content. More specifically, in the truncated representation theory, the representation $a=k$, that is $\langle k-1,0\rangle$, consists of the $U_q(\msu(2))\times U_q(\msu(2))$ representation $(k-2,0)$ only, while for $a=k-1$ we have $\langle k-2,0\rangle=(k-2,1)\oplus(k-3,0)$. The fact that the tower of states is restricted to $a=1,2,\ldots,k$ also meshes perfectly with the dispersion relation for these states (\eqref{dismag} in Appendix \ref{A2}) which has a built-in periodicity $a\to a+2k$ and symmetry $a\to 2k-a$:
\EQ{
\sin^2\Big(\frac{\xi E}{2g}\Big)-\xi^2\sin^2\Big(\frac p{2g}\Big)=
(1-\xi^2)\sin^2\Big(\frac{\pi a}{2k}\Big)\ .
\label{dismag2}
}

Although we will not investigate the representation theory of the quantum supergroup $U_q(\mh)$ when $q$ is a root of unity in any detail here, one can infer what happens from thinking about the bosonic sub-algebra $U_q(\msu(2))\times U_q(\msu(2))$. In Appendix \ref{a5} we review the way that the representations of $U_q(\msu(2))$ satisfy a truncated Clebsch-Gordon decomposition. The relevant representations are $V_j^{(+1)}$, which we label as $(2j)$ above,  {with $j\leq k/2-1$,} whose tensor product decomposition takes the form \eqref{ttp}. This implies that the particular short representations $\langle a-1,0\rangle$ of $U_q(\mh)$ that describe the magnons have the truncated tensor product, when $q=e^{i\pi/k}$, of the form
\EQ{
\langle a_1-1,0\rangle\, \widetilde{\otimes}\, \langle a_2-1,0\rangle = \bigoplus_{m=|a_1-a_2|}^{\text{min}(a_1+a_2-2,2k-a_1-a_2-2)}\, \{m,0\}\ ,
\label{ttp4}
}
where $\{m,0\}$ is a long representation. There is a similar expression for the solitons involving representations $\langle0,a-1\rangle$.
This decomposition meshes with the bootstrap
procedure of S-matrix theory (reviewed in Appendix \ref{a4}). The S-matrix for $a_1$ scattering with $a_2$ (either magnon or soliton) has two direct channel poles shown in Figure \ref{f8}. The pole I is physical if $a_1+a_2\leq k$ and the term $\{a_1+a_2-2,0\}$ becomes reducible (but indecomposable) implying that the bound state transforms in the short representation $\langle a_1+a_2-1,0\rangle$ as shown in Figure \ref{f8}. What is particularly noteworthy here is that the kinematical condition $a_1+a_2\leq k$ dovetails perfectly with the truncation of the Clebsch-Gordon decomposition. At the latter pole II, the term $\{|a_1-a_2|,0\}$ becomes reducible (but indecomposable) implying that the bound state transforms in the short representation $\langle |a_1-a_2|-1,0\rangle$ as shown in Figure \ref{f8} for $a_1\geq a_2$. There is a similar discussion for the solitons.

\section{Discussion}

In this work we have argued that the $q$-deformation of the string world-sheet S-matrix in $\text{AdS}_5\times S^5$ is described by an IRF, or RSOS, type S-matrix. The original ``vertex" form of the S-matrix is just a starting point for the vertex-to-IRF transformation. Unlike its vertex cousin, the new IRF S-matrix is manifestly unitary. It also satisfies all the S-matrix axioms familiar from a relativistic theory, albeit with a more complicated analytic structure. The bootstrap equations are discussed in Appendix \ref{a4}, including some strong checks that they mesh with the representation theory.

We found that in the relativistic limit, $g \rightarrow \infty$, the perturbative tree-level S-matrix of the Pohlmeyer-reduced theory is recovered in a particular semi-classical expansion. The details of this expansion clarify why the tree-level result satisfies a deformed classical Yang-Baxter equation. While we found agreement at the tree level, at one-loop there is still a discrepancy. This might be expected as the perturbative computations with which we are comparing assumed trivial boundary conditions. This is in contrast with the excitations whose scattering is described by the kink S-matrix. To find agreement at higher orders one would then need to incorporate the non-trivial boundary conditions and introduce the $\msu(2)$ spins $j$ and $l$ into the perturbative computation. This remains an open problem, however, progress in this direction has been made in \cite{Hollowood:2013oca}, in which an action was constructed with the required properties. %refrep

In the other limit of interest, $q \rightarrow 1$, we recover the string S-matrix. Again this limit is subtle and should be taken in such a way that preserves the symmetry between $j,l = 0$ and $j,l = \tfrac{k}{2}-1$, the end-points of the range of vacua. This can be done by just performing a shift in $j,l$ such the range is symmetric about $0$, see Eq.~\eqref{rdef}. The shifted vacuum labels $\tilde{\jmath}$ and $\tilde{l}$ then drop out of the S-matrix in the $q \to 1$ limit. It is an interesting open question as to whether the shifted vacuum labels have a physical meaning in the string limit or are just an artefact of the non-trivial vacuum structure of the interpolating theory. This could be the case if, for example, the vacua become degenerate as $q \rightarrow 1$. %refrep

Strong evidence that the quantum group restriction is the correct procedure to apply to the S-matrix in the present context could be obtained by studying the Thermodynamic Bethe Ansatz in the relativistic $g\to\infty$ limit.\footnote{We would like to thank the referee of this paper for suggesting this as a worthwhile analysis.} If the central charge of the UV theory could be extracted then this could be compared to the central charge of the UV CFT; namely, the $G/H$ gauge WZW model. Whilst this analysis has yet to be done for the string theory case, in the simpler context of the purely bosonic symmetric space $\CP^2$, where 
$G/H=\SU(2)/U(1)$, the calculation of the central charge from the TBA has been performed \cite{Hollowood:2010rv} and precise agreement was found. This is additional circumstantial evidence that the quantum group restriction is the correct paradigm. 

Finally, one of the key messages of this work is that the interpolating theory has a non-trivial vacuum structure with kinks playing the r\^ole of one-particle states. Therefore this should be respected by any fundamental off-shell (Lagrangian or otherwise) origin for the interpolating theory. In fact, for the Pohlmeyer reduced theory, whose Lagrangian description is known, the non-trivial vacuum structure arises  {as} a consequence of the presence of a WZW term in the bosonic sector of the theory~\cite{Hollowood:2013oca}.

\section*{Acknowledgements}

\noindent
BH is supported by the Emmy Noether Programme ``Gauge fields from Strings'' funded by the German Research Foundation (DFG). He would like to thank Gleb Arutyunov, Wellington Galleas, Arkady Tseytlin and Stijn van Tongeren for useful discussions.

\noindent
TJH is supported in part by the STFC grant ST/G000506/1 and further acknowledges support 
from the European Science Foundation (ESF) for the activity entitled ``Holographic Methods for Strongly Coupled Systems".

\noindent
JLM is supported in part by MINECO (FPA2011-22594), the Spanish Consolider-Ingenio 2010 Programme CPAN (CSD2007-00042), and FEDER. He would like to thank Manuel Asorey and Joaqu\'\i n S\'anchez Guill\'en for useful discussions.

\vspace{1cm}

\appendix
\appendixpage

\section{The Vertex-to-IRF Transformation}\label{A1}

The required $q$-deformed Clebsch-Gordan ($q$-CG) coefficients in our conventions are~\cite{Kirillov:1988,Ardonne:2010zu}
\EQ{
\left[\ARR{j+\frac12 & \frac12 & j\\
m\pm\frac12 & \pm\frac12 & m}\right]_q&=q^{\pm j/2-m/2}\sqrt{\frac{[j\pm m+1]}{[2j+1]}}\ ,\\
\left[\ARR{j-\frac12 & \frac12 & j\\
m\pm\frac12 & \pm\frac12 & m}\right]_q&=\pm q^{\mp(j+1)/2-m/2}\sqrt{\frac{[j\mp m]}{[2j+1]}}\ .
\label{qCG}
}
They correspond to the basis of $V_j$ given by $\ket{j,m}$  with $m=-j,-j+1,\ldots,j$, on which the action of the generators is
\EQ{
J_\pm \ket{j,m}= \sqrt{[j\mp m]\, [j\pm m+1]}\, \ket{j,m\pm1}\,,\qquad
H \ket{j,m}= 2m\ket{j,m}\,.
}
This basis is easily related to the one considered in~\eqref{repA} for $V_j^{(+1)}\equiv V_j$. The q-CG coefficients~\eqref{qCG} are consistent with the following definition of the co-product
\EQ{
\Delta(J_\pm)=J_\pm\otimes q^{-H/2} + q^{H/2}\otimes J_\pm\,,\qquad
\Delta(H)=H\otimes 1+1\otimes H\,.
}
When $q$ is a root of unity, say $q^{2k}=1$, notice that both {q-CG coefficients} are singular at $2j+1\in k{\mathbb Z}$. However, as explained in section~\ref{s2} {and Appendix~\ref{a5}}, in this case the S-matrix theory only involves kinks associated to the vacua labelled by the finite set $j\in\{0,\frac12,\ldots, \frac{k}2-1\}$, and $[2j+1]$ is always non-vanishing.

In~\eqref{mpsL}, consider the two-particle state
\EQ{
&\ket{\Phi_{j_1,j_2} (\theta_1)\Phi_{j_2,j_3}(\theta_2)}^{M_1}{}_{M_3}=\\
&\qquad\qquad\sum_{\{m_i=\pm\frac12\}}\left[\ARR{j_1 & \tfrac12 & j_2\\
M_1 & m_1 & M_2}\right]_q
\left[\ARR{j_2 & \frac12 & j_3\\
M_2 & m_2 & M_3}\right]_q\, \ket{\phi_{m_1}(\theta_1)\phi_{m_2}(\theta_2)}\ ,
}
where $M_2=M_3+m_2$ and
the sum is restricted to $M_1=M_3+m_1+m_2$.
It is an important part of our story that the S-matrix does not depend on the $j_z$ quantum numbers $M_1$ and $M_3$, subject to the fact that $M_3$ must be either $M_1\pm1$ or $M_1$.
As an example of works, let us consider the following set of states with $j_1=j+1$ and $j_3=j$, and for illustration we choose $M_3=j$:
\EQ{
&\ket{\Phi_{j+1,j+\frac12}(\theta_1)\Phi_{j+\frac12,j}(\theta_2)}^{j+1}{}_j=\ket{\phi_{\frac12}(\theta_1)\phi_{\frac12}(\theta_2)}\ ,\\[5pt]
&\ket{\Phi_{j+1,j+\frac12}(\theta_1)\Phi_{j+\frac12,j}(\theta_2)}^j{}_j=\frac{q^{-j-1/2}}{\sqrt{[2j+2]}}\Big(q\,\ket{\phi_{\frac12}(\theta_1)
\phi_{-\frac12}(\theta_2)}+\ket{\phi_{-\frac12}(\theta_1)
\phi_{\frac12}(\theta_2)}\Big)\ ,\\[5pt]
&\ket{\Phi_{j+1,j+\frac12}(\theta_1)\Phi_{j+\frac12,j}(\theta_2)}^{j-1}{}_j=q^{-2j}\sqrt{\frac{[2]}{[2j+2][2j+1]}}\,\ket{\phi_{-\frac12}(\theta_1)
\phi_{-\frac12}(\theta_2)}\ ,
}
Given the action of the S-matrix in the vertex picture \eqref{wea} one can easily verify, using the identity \eqref{sid}, that
\EQ{
\ket{\Phi_{j+1,j+\frac12}(\theta_1)\Phi_{j+\frac12,j}(\theta_2)}^{M_1}{}_{M_3}\longrightarrow S_I(\theta)\,\ket{\Phi_{j+1,j+\frac12}(\theta_1)\Phi_{j+\frac12,j}(\theta_2)}^{M_1}{}_{M_3}\ ,
}
independently of $M_1$ and $M_3$. This gives one of the elements in the first line of \eqref{web}. The other is given by taking $j_1=j-1$ in a similar way. To derive the elements in the second line it is sufficient to consider
\EQ{
&\ket{\Phi_{j,j+\frac12}(\theta_1)\Phi_{j+\frac12,j}(\theta_2)}^{j}{}_j=
\frac{q^{-2j-1/2}}{\sqrt{[2j+2][2j+1]}}
\ket{\phi_{\frac12}(\theta_1)\phi_{-\frac12}(\theta_2)}
-q^{1/2}\sqrt{\frac{[2j+1]}{[2j+2]}}\,\ket{\phi_{-\frac12}(\theta_1)\phi_{\frac12}(\theta_2)}
\ ,\\[5pt]
&\ket{\Phi_{j,j-\frac12}(\theta_1)\Phi_{j-\frac12,j}(\theta_2)}^{j}{}_j=-
q^{\frac12}\sqrt{\frac{[2j]}{[2j+1]}}\,
\ket{\phi_{\frac12}(\theta_1)\phi_{-\frac12}(\theta_2)}\ .
}
By using the action of the S-matrix in the vertex picture \eqref{wea} one can verify that
\EQ{
&\ket{\Phi_{j,j\pm\frac12}(\theta_1)\Phi_{j\pm\frac12,j}(\theta_2)}^j{}_j\longrightarrow\frac{\sqrt{[2j][2j+2]}}{[2j+1]}S_T(\theta)\ket{\Phi_{j,j\mp\frac12}(\theta_1)\Phi_{j\mp\frac12,j}(\theta_2)}^j{}_j\\[5pt]
& \qquad\qquad\quad~~~~~~~~~~~~~~~~~
+\frac{q^{2j+1}S_R^\mp(\theta)-q^{-2j-1}S_R^\pm(\theta)}{q^{2j+1}-q^{-2j-1}}
\ket{\Phi_{j,j\pm\frac12}(\theta_1)\Phi_{j\pm\frac12,j}(\theta_2)}^j{}_j\ .
}
This gives the second line of \eqref{web} once the $j_z$ quantum numbers are hidden.

The vertex-to-IRF transformation can also be used to derive the appropriate charge conjugation matrix for the kinks. Consider the one-particle states
\EQ{
& \ket{\Phi_{j,j+\frac12}(\theta)}^M{}_{M\mp \frac12} = \left[\ARR{ j & \frac12 & j+\frac12 \\ M & \pm \frac12 & M \mp \frac12}\right]_q \ket{\phi_{\pm \frac12}(\theta)} \ , \\ 
& \ket{\Phi_{j+\frac12,j}(\theta)}^M{}_{M\mp \frac12} = \left[\ARR{ j+\frac12 & \frac12 & j \\ M  & \pm \frac12 & M\mp\frac12}\right]_q \ket{\phi_{\pm \frac12}(\theta)} \ .}
It then immediately follows from the charge conjugation of $\ket{\phi_{\pm \frac12}(\theta)}$~\eqref{phicross} that
\EQ{
{\cal C}\ket{\Phi_{j+\frac12,j}}^{M\mp\frac12}{}_{M} =
-i\, \sqrt{\frac{[2j+2]}{[2j+1]}}\,\ket{\Phi_{j,j+\frac12}}^M{}_{M\mp\frac12}\,.
}
Therefore, since under the vertex-to-IRF transformation $\ket{\Phi_{j_1j_2}}^{M_1}{}_{M_2}\to \ket{K_{j_1j_2}}$, the charge conjugation matrix acting on the kinks is independent of the $j_z$ quantum numbers and is given by
\EQ{{\cal C}\ket{K_{ab}(\theta)} = e^{i\pi(b-a)}\sqrt{\frac{[2a+1]}{[2b+1]}}\ket{K_{ba}(\theta)} \ .}

\section{\texpdf{$\boldsymbol q$}{q}-Deformed World-sheet S-Matrix Kinematics}\label{A2}

The deformed theories that lie along the red lines in Figure \ref{f1} are non-relativistic. This is most apparent in the energy and momentum dispersion relation which takes the form
\EQ{
\sin^2\Big(\frac{\xi E}{4g}\Big)-\xi^2\sin^2\Big(\frac p{4g}\Big)=
(1-\xi^2)\sin^2\Big(\frac{\pi a}{2k}\Big)\ .
\label{dismag}
}
In the above $a=1,2,\ldots,k$ is an integer charge and 
$\xi$ is a parameter that lies between 0 and 1 and takes the value
\EQ{
\xi=\frac{2g\sin(\pi/k)}{\sqrt{1+4g^2\sin^2(\pi/k)}}\ .
}
Note that we have scaled the energy is a different way compared with \cite{Hoare:2012fc}.\footnote{In fact, $E_\text{here}=2E_\text{there}/\xi$ and $p_\text{here}=2p_\text{there}$.} The present scaling correctly gives the energy in both the string and relativistic limits, $k\to\infty$ and $g\to\infty$, respectively. Note that in \cite{Hoare:2012fc} the quantum group variables $U$ and $V$ \cite{Beisert:2011wq} that have a trivial group-like co-product on two-particle states are related to the energy and momentum in the following way:
\EQ{
U= e^{ip/4g}  \ , \qquad V = e^{i\xi E/4g}\ .
}

\begin{figure}[ht]
 \begin{minipage}[t]{.47\textwidth}
    \begin{center}
    \begin{tikzpicture}[scale=0.4]
\draw[decoration={brace,amplitude=0.5em},decorate,very thick] (4,11.8) -- (0,11.8);
\draw[decoration={brace,amplitude=0.5em},decorate,very thick] (12,11.8) -- (4,11.8);
\node at (2,10.5) (b1) {\bf soliton};
\node at (8,10.25) (b2) {\bf magnon};
\draw[-] (0,0) -- (12,0) -- (12,12) -- (0,12) -- (0,0);
\node at (0,-0.8) (a1) {$0$};
\node at (4,-0.8) (a2) {$\frac{2\pi a g}k$};
\node at (12,-0.8) (a1) {$2\pi g$};
\node at (7,-1.8) (a1) {$p$};
\node at (-1.7,7) (a1) {$E$};
\draw[densely dashed] (4,0) -- (4,12);
\draw[densely dashed] (-0.2,4.515) -- (4,4.515);
\node at (-1.2,4.515) (a3) {$\frac{2\pi a g}{\xi k}$};
\draw[very thick,densely dashed] plot[smooth] coordinates {(0., 2.01838)  (1., 2.2606)  (2., 2.86478)  (3., 3.65017)  (4., 
  4.51514)};
\draw[very thick] plot[smooth] coordinates   {(4., 4.51514)  (4.57143, 5.02511)  (5.14286, 5.53909)  (5.71429, 
  6.05263)  (6.28571, 6.56165)  (6.85714, 7.06206)  (7.42857, 
  7.54933)  (8., 8.01805)  (8.57143, 8.46154)  (9.14286, 
  8.87133)  (9.71429, 9.23672)  (10.2857, 9.54471)  (10.8571, 
  9.78052)  (11.4286, 9.92956)  (12., 9.98065)} ;
 \end{tikzpicture}
    \end{center}
  \end{minipage}
  \hfill
  \begin{minipage}[t]{.47\textwidth}
    \begin{center}
\begin{tikzpicture}[xscale=0.4,yscale=4.8]
\draw[-] (0,0) -- (12,0) -- (12,1) -- (0,1) -- (0,0);
\node at (0,-0.06) (a1) {$0$};
\node at (-0.7,1) (a1) {$1$};
\node at (12,-0.06) (a1) {$2\pi g$};
\node at (7,-0.15) (a1) {$p$};
\node at (-1.7,0.5) (a1) {$v$};
\node at (4,-0.07) (a2) {$\frac{2\pi a g}k$};
\draw[densely dashed] (4,0) -- (4,1);
\draw[densely dashed] (-0.2,0.885) -- (4,0.885);
\node at (-0.9,0.885) (a3) {$\xi$};
\draw[very thick,densely dashed] plot[smooth] coordinates  {(0., 0.)  (1., 0.458023)  (2., 0.71837)  (3., 0.836332)  (4., 
  0.885908)};
\draw[very thick] plot[smooth] coordinates  {(4., 0.885908)  (4.57143, 0.897379)  (5.14286, 0.900295)  (5.71429, 
  0.895896)  (6.28571, 0.884491)  (6.85714, 0.865645)  (7.42857, 
  0.83823)  (8., 0.800373)  (8.57143, 0.749386)  (9.14286, 
  0.681744)  (9.71429, 0.593329)  (10.2857, 0.480264)  (10.8571, 
  0.340725)  (11.4286, 0.177535)  (12., 0)};
 \end{tikzpicture}
    \end{center}
  \end{minipage}
\caption{\small On the left the energy as a function of momentum and on the right the velocity for the case $k=6$, $a=2$ and $g=6/\pi$ with $0\leq p\leq 2\pi g$. The dotted part of the curves represent the soliton branch $|p|<2\pi a g/k$ and the continuous part the magnon branch with $|p|>2\pi a g/k$.
}
\label{f4}
\end{figure}
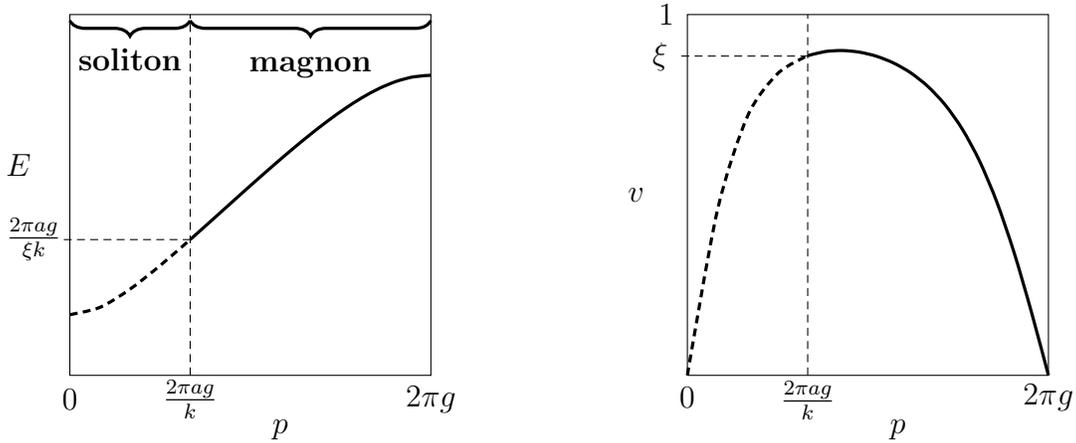

The momentum is restricted to the interval $|p|\leq 2\pi g$ and the energy is plotted in Figure \ref{f4} for positive momenta. The velocity of a state is identified with the group velocity of a wave packet, $v=\partial E/\partial p$ and so this rises to a maximum and then goes to zero as $p$ approaches $2\pi g$, also shown in Figure \ref{f4}. Note the counter-intuitive fact that there are two different states of the particle with a given velocity distinguished by having different momenta. The maximum velocity is less than 1, the relativistic speed of light. Also shown in the Figure are the two distinguished branches corresponding to $|p|\gtrless 2\pi a g/k$. The small/large momentum branch was called the soliton/magnon branch in \cite{Hoare:2012fc}. The states in the two branches carry different quantum numbers under the $\mpsu(2|2)^{\times 2}$ global quantum group symmetry, namely $\langle0,a-1\rangle$ and $\langle a-1,0\rangle$, respectively. The state with $a=1$ is common to both branches. Note that the special point $|p|=2\pi ag/k$ corresponds to $E=2\pi ag/(\xi k)$ and so is precisely the point where the states become marginally unstable to decay to $a$ copies of the basic state $a=1$. Another way to think about the dividing line between the two branches is that it occurs at the specific value of the velocity $v=\xi$.

The relativistic limit corresponds to $g\to\infty$ with fixed $k$ in which case we recover the familiar relativistic dispersion relation
\EQ{
E^2-p^2=\frac{4\sin^2(\pi a/2k)}{\sin^2(\pi/k)}\ ,
}
which identifies the mass as
\EQ{
M = \frac{2\sin(\pi a/2k)}{\sin(\pi/k)}\ .
\label{mass}}
In the string limit, where $k\to\infty$ with fixed $g$, we find the familiar dyonic magnon dispersion relation
\EQ{
E^2=a^2+16g^2\sin^2\Big(\frac p{4g}\Big) .
}

There are three sets of alternative kinematic variables that are useful in the S-matrix theory:

{\bf (i)} Firstly, the pair $x^\pm$, which are related to energy and momentum via
\EQ{
e^{ip/2g}=q^{-a}\frac{x^++\xi}{x^-+\xi}=q^a\frac{\frac1{x^-}+\xi}{\frac1{x^+}+\xi}\ ,\qquad
e^{i\xi E/2g}=q^{-a}\frac{\xi x^++1}{\xi x^-+1}=q^a\frac{\frac\xi{x^-}+1}{\frac\xi{x^+}+1}\ ,
\label{jwwapp}
}
where $q=\exp(i\pi/k)$. The pair then satisfy the dispersion relation
\EQ{
q^{-a}\Big(x^++\frac1{x^+}+\xi+\frac1\xi\Big)=q^a\Big(x^-+\frac1{x^-}+\xi+\frac1\xi\Big)\ .
\label{p11}
}
Note that physical states with real energy and momentum satisfy the reality condition $(x^\pm)^*=x^\mp$. The origin of these variables goes back to the original construction of \cite{Beisert:2008tw} where the fundamental states were labelled by parameters $x^\pm$, satisfying a constraint, which is just the shortening condition for the $4$-dimensional representation of $\msu(2|2)$. In \cite{Beisert:2011wq} a rescaling and shift of $x^\pm$ was found for which the constraint equation becomes the more appealing expression above for $a=1$.
For the bound states transforming in the short representations $\langle a-1,0 \rangle$ or $\langle 0,a-1 \rangle$ the shortening condition is then \eqref{p11} above \cite{deLeeuw:2011jr,Hoare:2011wr,Hoare:2012fc}.

{\bf (ii)} The pseudo rapidity $\nu$. If we define the map $x(\nu)$ via
\EQ{
x+\frac1x+\xi+\frac1\xi=\Big(\frac1\xi-\xi\Big)e^{2\nu}\ ,
}
then this variable determines the pair $x^\pm$ via
\EQ{
x^\pm=x\Big(\nu\pm\frac{i\pi a}{2k}\Big)\ .
}
The pseudo rapidity $\nu$ plays an important role in the bootstrap equations because the bound state of $a$ basic particles ($a=1$) transforming in representations $\langle a-1,0\rangle$ corresponds the set of conditions
\EQ{
x_1^+=x_2^-\ ,\quad x_2^+=x_3^-\ , \ldots\ ,\quad x_{a-1}^+=x_a^-\ ,
} 
The bound state has kinematic variables $x^+=x_a^+$ and $x^-=x^-_1$. In terms of the pseudo rapidity, if $\nu$ is the pseudo rapidity of the bound state then its constituents have
\EQ{
\nu_j=\nu-\frac{i\pi}{2k}(a+1-2j)\ ,
}
$j=1,2,\ldots,a$. Correspondingly, if the bound state transforms in the representation $\langle 0,a-1\rangle$ then the constituents have pseudo rapidities $\nu_j=\nu+\frac{i\pi}{2k}(a+1-2j)$.

In addition, in the relativistic limit $\nu$ becomes the 
conventional rapidity in the relativistic limit. Note that  $\nu$ is  a re-scaled version of the variable $u$ considered in~\cite{Hoare:2011wr}: $\nu=\pi u/k$. On the magnon branch for physical values of the parameters (real energy and momentum) $\nu$ is real while on the soliton branch $\nu-i\pi/2$ is real, so that $e^{2\nu}\gtrless0$, respectively.

{\bf (iii)} The pair $z^\pm$ which are related to the $x^\pm$ by the following fractional linear transformation
\EQ{
x^\pm=\frac{z^\pm+\sigma}{z^\pm-\sigma}\ .
\label{huu}
}
The pair $z^\pm$ satisfy the dispersion relation
\EQ{
q^{2a}=\frac{(\sigma z^+)^2-1}{\sigma^2- (z^+)^2}\cdot\frac{\sigma^2- (z^-)^2}{(\sigma z^-)^2-1}\ ,
\label{zdr}
}
where we have defined 
\EQ{
\sigma=\sqrt{\frac{1+\xi}{1-\xi}}=2g\sin(\pi/k)+\sqrt{1+4g^2\sin^2(\pi/k)}\ .
}
These variables are particularly nice because if we write
\EQ{
z^\pm=e^{\theta\pm i\alpha}\ ,
\label{zdf}
}
so that for physical states (real energy and momentum) $\alpha$ and $\theta$ are real and so $(z^\pm)^*=z^\mp$, then if we solve for $\alpha=\alpha(\theta)$ then $\theta$ is the rapidity: $v=\tanh\theta$.
In addition, the soliton and magnon branches are then distinguished by  $\alpha(\theta)\gtrless0$, respectively.
In Figure \ref{f6} we plot $z^+$ in the complex plane to show the two distinct branches.
\FIG{
\begin{tikzpicture}[scale=1]
\draw[->] (-0.9,0) -- (5,0);
\draw[->] (0,-2.9) -- (0,1.5);
\node at (4.6,1.3) (a1) {$e^{\theta+i\alpha}$};
\draw (4,1.6) -- (4,0.9) -- (5,0.9);
\node at (2,1.7) (a2) {soliton};
\node at (5.2,-1.8) (a3) {magnon};
\node at (2,-0.4) (a4) {$p=0$};
\node at (2.3,-1.3) (a5) {$p=2\pi g$};
\draw[->] (a2) -- (2.4,1);
\draw[->] (a3) -- (4.2,-1.4);
\draw[->] (a4) -- (1,0.43);
\draw[->] (a5) -- (0.65,-0.84);
\draw[very thick,densely dashed] plot[smooth] coordinates   {(0.890302, 0.455371)  (1.1523, 0.579295)  (1.48061, 
  0.705691)  (1.87071, 0.811159)  (2.31329, 0.866161)  (2.7918, 
  0.837788)  (3.27687, 0.694244)  (3.72181, 
  0.414082)  (4.06568, 0) };
\draw plot[smooth] coordinates   {(4.06568, 0)  (4.24904, -0.509928)  
(4.2383, -1.04567)  (4.04274, -1.52798)  (3.7097, -1.89743)  
(3.30201, -2.12981)  (2.87549, -2.23207)  (2.46794, -2.22842)  
(2.09929, -2.14748)  (1.77658, -2.01503)  (1.49942, -1.85152)  
(1.26388, -1.67217)  (1.06491, -1.4881)  (0.897603, -1.30742)  
(0.757653, -1.13616)  (0.641468, -0.97868)  (0.545992, -0.837791)} ;
 \end{tikzpicture}
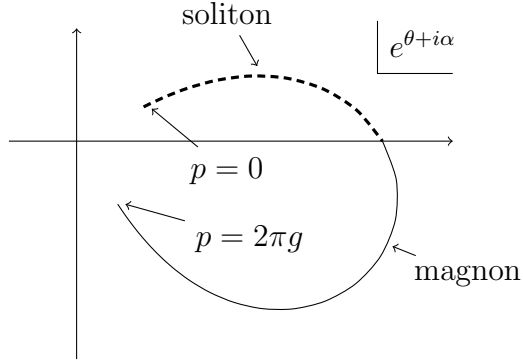
\caption{\small The momentum dependence of $z^+=e^{\theta+i\alpha}$ showing the soliton and magnon branches distinguished by the sign of $\alpha$.}
\label{f6}
}

The crossing symmetry transformation is the same as in a relativistic theory $\theta\to\theta+i\pi$. 
It follows that 
\EQ{
\alpha(i\pi+\theta,a)=\alpha(\theta,a)\ ,
}
and also from \eqref{huu} that
\EQ{
x^\pm(i\pi+\theta)=\frac1{x^\pm(\theta)}\ ,
}
which is the known transformation \cite{Beisert:2008tw,Beisert:2011wq,Hoare:2011wr}. The function $\alpha$ also satisfies
\EQ{
\alpha(\theta,a)=\alpha(-\theta,a)\ .
}

The relativistic limit is obtained by taking $g\to\infty$. In this limit, $q^a=e^{2i\alpha}$, or $\alpha=\pi a/2k$,\footnote{Note that here $x^\pm+1 \sim e^{+\theta}$. If we consider the $q$-deformed dressing phase (denoted $\sigma$ in \cite{Hoare:2011wr}) then it is known \cite{Hoare:2011wr} that taking the $+$ sign and requiring the usual relativistic crossing relation in the $g \rightarrow \infty$ limit implies that the bound states transform in the $\langle 0,a \rangle$ representation in this same limit. In \cite{Hoare:2012fc} it was shown that this is indeed the case and therefore the $+$ sign here is consistent.}
\EQ{
x^\pm=-1-\frac{e^{\theta\pm i\alpha}}{2g\sin(\pi/k)}+{\cal O}(g^{-2})\ .
}
If we then take the semi-classical limit $k\to\infty$ with $\alpha$ fixed of \eqref{mass}, we find that the states have a mass $M=2k/\pi\sin\alpha$ which matches the masses of the solitons in the generalized sine-Gordon theory in \cite{Hollowood:2011fq,Hollowood:2011fm} with the identification $\alpha=q$. (Note that $q$ here is not identified with the $q$-deformation parameter of the quantum group).

\noindent{\bf The Mirror Theory}\nopagebreak

This is obtained by transforming $p\to-iE$ and $E\to-ip$ and so the dispersion relation takes the form
\EQ{
\xi^2\sinh^2\Big(\frac{E}{4g}\Big)-\sinh^2\Big(\frac{\xi p}{4g}\Big)=
(1-\xi^2)\sin^2\Big(\frac{\pi a}{2k}\Big)\ .
\label{dismagmir}
}
The variables $x^\pm$ are related to the energy and momentum via
\EQ{
e^{E/2g}=q^{-a}\frac{x^++\xi}{x^-+\xi}=q^a\frac{\frac1{x^-}+\xi}{\frac1{x^+}+\xi}\ ,\qquad
e^{\xi p/2g}=q^{-a}\frac{\xi x^++1}{\xi x^-+1}=q^a\frac{\frac\xi{x^-}+1}{\frac\xi{x^+}+1}\ ,
\label{jwwmir}
}
and they satisfy the same dispersion relation \eqref{p11} but now the physical reality condition is $x^{+*}=1/x^-$. In
 this case there is only a single branch of physical states corresponding to pseudo rapidity $\nu\in{\mathbb R}$.
 
The mapping to the rapidity is given by 
\EQ{
x^\pm=\frac{z^\pm+i\sigma}{z^\pm-i\sigma}\ .
\label{huumir}
}
with $z^\pm=e^{\theta\pm i\alpha}$ and the dispersion relation \eqref{zdr} is changed to
\EQ{
q^{2a}=\frac{(z^+\sigma)^2+1}{\sigma^2+(z^+)^2}\cdot\frac{\sigma^2+ (z^-)^2}{(\sigma z^-)^2+1}\ .
\label{zdr2mir}
}
In this case, $\alpha>0$ for all momenta.

\section{The \texpdf{``Free''}{''Free''} S-Matrix}\label{A3}

The $q$-deformed vertex and kink S-matrices, \eqref{jjs} and \eqref{web2} respectively, depend on two couplings $g$ and $k$. As discussed in section \ref{slim}, in the $k \rightarrow \infty$ and $g \rightarrow \infty$ limits we recover the light-cone string and the Pohlmeyer-reduced theory S-matrices. From the perspective of two-dimensional field theories the standard perturbative expansions are in powers of $1/g$ and $1/k$ respectively. Motivated by this we briefly discuss the ``free'' theory limit ($g \rightarrow \infty$ and $k \rightarrow \infty$) of the $q$-deformed vertex and kink S-matrices. In both cases these limits commute.

In this ``free'' limit the vertex S-matrix is just the permutation operator
\EQ{
&\ket{\phi_m\phi_m} \longrightarrow \ket{\phi_m\phi_m}\ , \qquad~~  \ket{\psi_m\psi_m} \longrightarrow -\ket{\psi_m\psi_m}\ , \\
&\ket{\phi_{\pm\frac12}\phi_{\mp\frac12}} \longrightarrow \ket{\phi_{\mp\frac12}\phi_{\pm\frac12}} \ , \quad  \ket{\psi_{\pm\frac12}\psi_{\mp\frac12}} \longrightarrow -\ket{\psi_{\mp\frac12}\psi_{\pm\frac12}} \ ,\\
&\ket{\phi_m\psi_n} \longrightarrow \ket{\psi_n\phi_m}\ , \qquad ~~~ \ket{\psi_m\phi_n} \longrightarrow \ket{\phi_n\psi_m}\ .
\label{jjsfree}
}
The kink S-matrix however has a more complicated structure:
\EQ{
 \ket{K^{l,l}_{j\mp\frac12,j}K^{l,l}_{j,j\pm\frac12}} \longrightarrow & \ \ket{K^{l,l}_{j\mp\frac12,j}K^{l,l}_{j,j\pm\frac12}}\ ,\\
 \ket{K_{j,j}^{l\mp\frac12,l}K_{j,j}^{l,l\pm\frac12}} \longrightarrow & \ - \ket{K_{j,j}^{l\mp\frac12,l}K_{j,j}^{l,l\pm\frac12}}\ ,\\
 \ket{K^{l,l}_{j,j\pm\frac12}K^{l,l}_{j\pm\frac12,j}} \longrightarrow & \ \mp\frac{1}{2j+1} \ket{K^{l,l}_{j,j\pm\frac12}K^{l,l}_{j\pm\frac12,j}}\\ & \ +
\frac{\sqrt{2j(2j+2)}}{2j+1} \ket{K^{l,l}_{j,j\mp\frac12}K^{l,l}_{j\mp\frac12,j}}\ ,\\
 \ket{K_{j,j}^{l,l\pm\frac12}K_{j,j}^{l\pm\frac12,l}} \longrightarrow & \ \pm \frac{1}{2l+1} \ket{K_{j,j}^{l,l\pm\frac12}K_{j,j}^{l\pm\frac12,l}}\\ & \ -
\frac{\sqrt{2l(2l+2)}}{2l+1} \ket{K_{j,j}^{l,l\mp\frac12}K_{j,j}^{l\mp\frac12,l}}
\ ,\\
 \ket{K_{j\pm\frac12,j}^{l,l}K_{j,j}^{l,l\pm\frac12}} \longrightarrow & \
\ket{K_{j\pm\frac12,j\pm\frac12}^{l,l\pm\frac12}K_{j\pm\frac12,j}^{l\pm\frac12,l\pm\frac12}}\ , \\
 \ket{K_{j,j}^{l\pm\frac12,l}K_{j,j\pm\frac12}^{l,l}} \longrightarrow & \
\ket{K_{j,j\pm\frac12}^{l\pm\frac12,l\pm\frac12}K_{j\pm\frac12,j\pm\frac12}^{l\pm\frac12,l}}\ ,
\label{webexfree}
}
Taking the $(j,l)\rightarrow \infty$ limit {as explained in Sec.~\ref{slim}} we recover the permutation operator (i.e. agreement with the vertex S-matrix). This confirms that it is only in this limit that we have a good semi-classical interpretation.

\section{The Bootstrap}\label{a4}

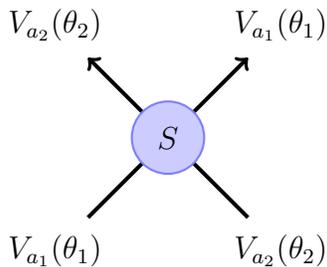
\begin{figure}
\begin{center}
\begin{tikzpicture} [line width=1.5pt,inner sep=2mm,
place/.style={circle,draw=blue!50,fill=blue!20,thick}]
\begin{pgfonlayer}{foreground layer}
\node at (1.5,1.5) [place] (sm) {$S$}; 
\end{pgfonlayer}
\node at (0,0) (i1) {$V_{a_1}(\theta_1)$};
\node at (3,0) (i2) {$V_{a_2}(\theta_2)$};
\node at (0,3) (i3) {$V_{a_2}(\theta_2)$};
\node at (3,3) (i4) {$V_{a_1}(\theta_1)$};
\draw[->] (i1) -- (i4);
\draw[->] (i2) -- (i3);
\end{tikzpicture}
\caption{\small The basic 2-body S-matrix elements that intertwine a tensor product of particle Hilbert spaces.}
\label{f1a}
\end{center}
\end{figure}

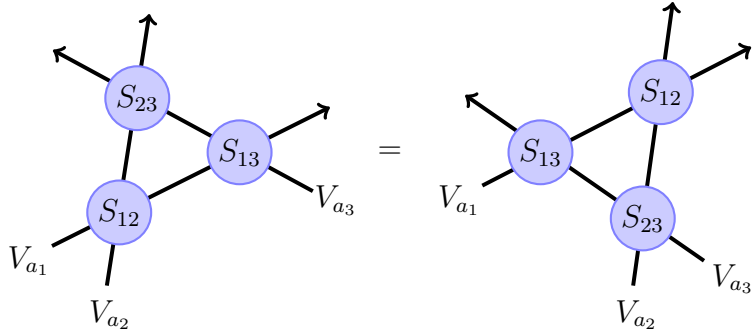
\begin{figure}
\begin{center}
\begin{tikzpicture} [scale=0.8,line width=1.5pt,inner sep=2mm,
place/.style={circle,draw=blue!50,fill=blue!20,thick}]
\begin{pgfonlayer}{top layer}
\node at (1.5,1.5) (pm1) {$S_{12}$}; 
\node at (1.8,3.4) (pm2) {$S_{23}$}; 
\node at (3.5,2.5) (m3) {$S_{13}$}; 
\node at (8.5,2.5) (m4) {$S_{13}$}; 
\node at (10.2,1.4) (m5) {$S_{23}$}; 
\node at (10.5,3.5) (m6) {$S_{12}$}; 
\end{pgfonlayer}
\begin{pgfonlayer}{foreground layer}
\node at (1.5,1.5) [place] (sm1) {\phantom{i}}; 
\node at (1.8,3.4) [place] (sm2) {\phantom{i}}; 
\node at (3.5,2.5) [place] (sm3) {\phantom{i}}; 
\node at (8.5,2.5) [place] (sm4) {\phantom{i}}; 
\node at (10.2,1.4) [place] (sm5) {\phantom{i}}; 
\node at (10.5,3.5) [place] (sm6) {\phantom{i}}; 
\end{pgfonlayer}
\node at (6,2.5) {$=$};
\node at (10,0) (i1) {};
\node at (10.1,-0.2) {$V_{a_2}$};
\draw[->] (i1) -- (10.7,5);
\node at (11.7,0.4) {$V_{a_3}$};
\node at (11.5,0.5) (i2) {};
\draw[->] (i2) -- (7.25,3.45);
\node at (7.25,1.8) (i3) {};
\node at (7.15,1.7) {$V_{a_1}$};
\draw[->] (i3) -- (12,4.25) ;
\node at (5,1.7) (i4) {};
\node at (5.1,1.7) {$V_{a_3}$};
\draw[->] (i4) -- (0.4,4.2);
\node at (0.1,0.8) (i5) {};
\node at (0,0.7) {$V_{a_1}$};
\draw[->] (i5) -- (5,3.25);
\node at (1.25,0) (i6) {$$};
\node at (1.35,-0.2) {$V_{a_2}$};
\draw[->] (i6) -- (2,4.8);
\end{tikzpicture}
\caption{\small 
The Yang-Baxter equation follows from the locality of interactions for widely separated wave packets and the fact that higher spin conserved charges can be used to shift the trajectories of the wavepackets without affecting the S-matrix.}
\label{f2a}
\end{center}
\end{figure}

As in a relativistic theory, integrability ensures that the S-matrix is factorizable. This means that if there are $n$ incoming particles with momenta $\{p_i\}$ then there are $n$ outgoing particles with momenta $\{p_i\}$: momenta are individually conserved. The locality of interactions, along with the fact that the centres of the $n$ incoming particles---or more properly wavepackets---can be moved relative to each other at will using transformations generated by the higher spin conserved charges of the integrable theory, means that the $n$-body S-matrix factorizes into pair-wise scatterings. The 2-body S-matrix elements can be thought of as intertwiners, or maps, between tensor products of vector spaces, as illustrated in Figure~\ref{f1a},
\EQ{
S(\theta_{1},\theta_2):\quad V_{a_1}(\theta_1)\otimes V_{a_2}(\theta_2)\longrightarrow V_{a_2}(\theta_2)\otimes V_{a_1}(\theta_1)\ .
}
\noindent Note that they interchange the two particles. A process involving more particles 
is then built up out of these basic elements.
For example for the scattering of three particles
\EQ{
S(\theta_1,\theta_2,\theta_3)=
S_{12}(\theta_1,\theta_2)S_{13}(\theta_1,\theta_3)S_{23}(\theta_2,\theta_3)\ .
\label{sws}
}
In this context in \eqref{sws} the subscripts, on say $S_{12}$ for example, are redundant, but for future use they just remind us which factors in the tensor product the element acts on. Consistency between the different ways to factorize the $n$-body S-matrix elements leads to the Yang-Baxter equation which is illustrated in Figure~\ref{f2a}
\EQ{
S_{12}(\theta_1,\theta_2)S_{13}(\theta_1,\theta_3)S_{23}(\theta_2,\theta_3)=
S_{23}(\theta_2,\theta_3)S_{13}(\theta_1,\theta_3)S_{12}(\theta_1,\theta_2)\ .
\label{ybe}
}

{A rather} non-trivial aspect of S-matrix theory is the analytic structure and its explanation in terms of bound states and anomalous thresholds. The exchange of stable bound states in either the $s$- or $t$-channels gives simple poles of the S-matrix under a particular analytic continuation of the momenta of the incoming particles. 

\FIG{
\begin{tikzpicture} [scale=0.7,line width=1.5pt,inner sep=2mm,
place/.style={circle,draw=blue!50,fill=blue!20,thick},proj/.style={circle,draw=red!50,fill=red!20,thick}]
\begin{pgfonlayer}{foreground layer}
\node at (1,0.8) [proj] (p1) {};
\node at (1,4.2) [proj] (p2) {};
\node at (-0.9,-0.4) (i1) {$V_{a_1}(\rho_1(\theta))$};
\node at (2.9,-0.4) (i2) {$V_{a_2}(\rho_2(\theta))$};
\node at (-0.9,5.6) (i3) {$V_{a_2}(\rho_2(\theta))$};
\node at (2.9,5.6) (i4) {$V_{a_1}(\rho_1(\theta))$};
\node at (1,2.5) (l2) {$V_b(\theta)$};
\end{pgfonlayer}
\draw[-] (i1) -- (p1);
\draw[-] (i2) -- (p1);
\draw[<-] (i3) -- (p2);
\draw[<-] (i4) -- (p2);
\draw[red] (p2) -- (l2);
\draw[red] (l2) -- (p1);
\end{tikzpicture}
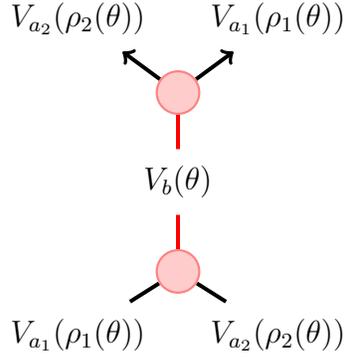
\caption{\small The S-matrix in the vicinity of a bound state pole. The bound state space $V_b(\theta)$ corresponds to the pre-image of the residue $\mR$ on the tensor product $V_{a_1}(\rho_1(\theta))\otimes V_{a_2}(\rho_2(\theta))$.}
\label{f4a}
}
The position of the bound-state poles must mesh with the representation theory of the symmetry algebra $U_q(\mh)$.\footnote{In the world-sheet case the symmetry algebra consists of a triple extension of two copies of the Lie superalgebra $\mathfrak{psu}(2|2)$ \cite{Beisert:2008tw}.}
For generic values of the rapidities, the representation on the product space $V_{a_1}\otimes V_{a_2}$ is irreducible.\footnote{In the following for brevity we often do not indicate the rapidity of the representation.} However, when the rapidities are analytically continued the S-matrix can exhibit a pole signalling the existence of a bound state. At the specific values of the incoming rapidities the product representation in general becomes reducible.

The conditions under which this happens are discussed in detail in \cite{Hoare:2012fc}. The momenta of the incoming particles must be analytically continued in a specific way; namely, $p_1=\tilde p_1+ir$ and $p_2=\tilde p_2-ir$, where particle 1 is coming in from the left so that the velocity $v_1>v_2$, and the imaginary part $r$ is positive. In a relativistic theory kinematics would require $\tilde p_1=\tilde p_2$, however in the non-relativistic setting this is not necessarily true. In terms of the rapidities let us say that the bound state occurs when
\EQ{
\theta_1=\rho_1(\theta)\ ,\qquad \theta_2=\rho_2(\theta)\ ,
\label{bsr}
}
where $\theta$ is identified as the rapidity of the bound state. 
At the pole the product representation must become reducible and contain the bound state representation $V_b$ as a component. That is at the specific rapidities \eqref{bsr}, there is a ``decomposition''
\EQ{
V_{a_1}(\rho_1(\theta))\otimes V_{a_2}(\rho_2(\theta))=V_b(\theta)\oplus V_b^\perp\ .
}
The assignment of $b=b(a_1,a_2)$ defines the ``fusion rule" for this process. In general for given initial states there will be more than one bound-state pole and therefore more than one possible fusion. If the residue of the S-matrix at the pole is defined as 
$\mR$ then we require that the other component 
$V^\perp_b$ lies in the kernel of the residue, that is
\EQ{
\mR:\; V_b^\perp\longrightarrow0\ .
\label{ipm}
}
For consistency with the symmetry algebra, it must be that $V_b^\perp$ is an invariant subspace with respect to the symmetry generators acting on the tensor product $V_{a_1}\otimes V_{a_2}$,
\EQ{
\Delta(u):\quad V_b^\perp\longrightarrow V_b^\perp\ ,\qquad u\in U_q(\mh)\ .
}
Therefore $V_b^\perp$ carries a representation of $\mh$, which is known as the ``sub-representation". Correspondingly, the bound state space $V_b$ is generally not an invariant subspace; rather
\EQ{
\Delta(u):\quad V_b\longrightarrow V_b\oplus V_b^\perp\ ,\qquad u\in U_q(\mh)\ .
}
It is in this sense that the product representation at the bound-state pole is reducible but indecomposable. The bound state space carries another representation, the factor representation, in the form of the quotient
\EQ{
V_\text{factor}=\coset{V_{b}\oplus V_b^\perp}{V_b^\perp}\ .
}
In all the cases that we consider this factor representation is irreducible.
If we introduce projectors $\mP$ and $\mP^\perp$ onto the subspace $V_b$ and $V_b^\perp$, respectively, then the action of the generators on the factor and sub-representations acting on $V_{a_1}\otimes V_{a_2}$ are given by\footnote{The following formulae are easier to digest in an explicit basis for $V_{a_1}\otimes V_{a_2}$ of the form $\MAT{V_b\\ V_b^\perp}$. In this basis,
\EQ{
\Delta(u)=\MAT{a & 0\\ b & c}\ ,\qquad\Delta_\text{factor}(u)=\MAT{a&0\\0&0}\ ,\qquad
\Delta_\text{sub}(u)=\MAT{0&0\\0&c}\ .}
Note that it is the fact that $b\neq0$ that makes the representation indecomposable.}
\EQ{
\Delta_\text{factor}(u)\equiv\mP\Delta(u)\ ,\qquad\Delta_\text{sub}(u)\equiv\Delta(u)\mP^\perp\ ,
}
for $u\in U_q(\mh)$. 

The fact that $b$ can appear as a bound state of $a_1$ and $a_2$ means that the S-matrix elements of $b$ with other states, say $a_3$, can be written in terms of those of $a_1$ and $a_2$. This is the essence of the bootstrap, or fusion, programme. The S-matrix element of the particle $a_3$ with rapidity $\theta_3$ and the bound-state with rapidity $\theta$ is written concretely as\footnote{It is important that the factor representation acting on $V_b$ is an irreducible representation of $\mh$. If it were not then the bound state S-matrix is not uniquely determined since it can be pre- and post-multiplied by $\sum_jr_j\mP^{(j)}_{12}$ and $\sum_jr_j^{-1}\mP^{(j)}_{12}$ where the $\mP^{(j)}$ are the projectors into the irreducible components and the numbers $r_j$ are arbitrary, and still satisfy the Yang-Baxter equation and commute with the generators of the symmetry \cite{Karowski:1978ps}.}
\EQ{
S_{12,3}(\theta,\theta_3)=\mP_{12} S_{13}(\rho_1(\theta),\theta_3)S_{23}(\rho_2(\theta),\theta_3)\ ,
\label{bfs}
}
where $V_b$ is to be thought of as a sub-space of $V_{a_1}\otimes V_{a_2}$ and the whole expression acts on the tensor product as
\EQ{
S_{12,3}(\theta,\theta_3):\qquad V_{a_1}\otimes V_{a_2}\otimes V_{a_3}\longrightarrow
V_{a_3}\otimes V_{a_1}\otimes V_{a_2}\ ,
}
but more specifically, as we show below, as
\EQ{
S_{12,3}(\theta,\theta_3)&:\qquad V_{b}\otimes V_{a_3}\longrightarrow
V_{a_3}\otimes V_{b}\ ,\\
S_{12,3}(\theta,\theta_3)&:\qquad V_{b}^\perp\otimes V_{a_3}\longrightarrow0\ .
\label{wer}
}

The first property \eqref{wer} follows trivially. The second follows from
the identity
\EQ{
S_{12,3}(\theta,\theta_3)=S_{12,3}(\theta,\theta_3)\mP_{12}\ ,
\label{ide}
}
which is proved as follows. First of all, we define an inverse for $\mR$ on $V_b$, $\mP\widetilde\mR\mR=\mP$, and then using the residue of the Yang-Baxter equation \eqref{ybe} evaluated at the pole,
\EQ{
\mR_{12}S_{13}(\rho_1(\theta),\theta_3)S_{23}(\rho_2(\theta),\theta_3)=
S_{23}(\rho_2(\theta),\theta_3)S_{13}(\rho_1(\theta),\theta_3)\mR_{12}\ ,
\label{rybe}
}
we have for the left-hand side of \eqref{ide}
\EQ{
\text{LHS}&=\mP_{12}\widetilde\mR_{12}\mR_{12} S_{13}(\rho_1(\theta),\theta_3)S_{23}(\rho_2(\theta),\theta_3)\\ &=\mP_{12}\widetilde\mR_{12}S_{23}(\rho_2(\theta),\theta_3)S_{13}(\rho_1(\theta),\theta_3)\mR_{12}
\\ &=\mP_{12}\widetilde\mR_{12}S_{23}(\rho_2(\theta),\theta_3)S_{13}(\rho_1(\theta),\theta_3)\mR_{12}\mP_{12}=\text{RHS}\ ,
}
where in the last line we used the fact that $\mR=\mR\mP$ and then applied the Yang-Baxter equation \eqref{rybe} again.
This identity implies that 
\EQ{
S_{12,3}(\theta,\theta_3)\mP_{12}^\perp=0\ .
}
which shows explicitly that the fused S-matrix vanishes on the subspace $V_b^\perp$ as claimed in \eqref{wer}.

It is simple to show using \eqref{ide}, that the fused S-matrix commutes with the action of the symmetry $U_q(\mh)$ on $V_b\otimes V_{a_3}\subset V_{a_1}\otimes V_{a_2}\otimes V_{a_3}$. Given that the action of the symmetry is given by
\EQ{
\mP_{12}\Delta(u):\quad V_{b}\otimes V_{a_3}
\longrightarrow V_{b}\otimes V_{a_3}\ ,
}
for $u\in U_q(\mh)$, we have
\EQ{
\big[\mP_{12}\Delta(u),S_{12,3}(\theta,\theta_3)\big]=0\ .
}

\FIG{
\begin{tikzpicture} [scale=0.7,line width=1.5pt,inner sep=2mm,
place/.style={circle,draw=blue!50,fill=blue!20,thick},proj/.style={circle,draw=red!50,fill=red!20,thick}]
\begin{pgfonlayer}{top layer}
\node at (3,3.6) (sm20) {$\mP$}; 
\node at (9.5,1.7) (sm21) {$\mP$};
\node at (1.5,2.3)  (m1) {$S_{13}$}; 
\node at (3.5,1.5) (m3) {$S_{23}$}; 
\node at (10.2,3.4) (m10) {$S_{12,3}$}; 
\end{pgfonlayer}
\begin{pgfonlayer}{foreground layer}
\node at (1.5,2.3) [place] (sm1) {\phantom{i}}; 
\node at (3,3.6) [proj] (sm2) {\phantom{i}}; 
\node at (3.5,1.5) [place] (sm3) {\phantom{i}}; 
\node at (10.2,3.4) [place] (sm10) {\phantom{i}}; 
\node at (9.5,1.7) [proj] (sm11) {\phantom{i}};
\node at (6.5,2.5) {$=$};
\end{pgfonlayer}
\node at (7.8,0.2) (k1) {$V_{a_1}$};
\draw[-] (k1) -- (sm11);
\node at (10.3,-0.2) (k2) {$V_{a_2}$};
\draw[-]  (k2) -- (sm11);
\node at (12.7,2.3) (k8) {$V_{a_3}$};
\draw[->] (k8) -- (8.2,4.2);
\node at (11.1,5.5) (k3) {$V_b$};
\draw[->,red] (sm11) -- (k3);
\node at (3.9,5.5) (k4) {$V_b$};
\draw[->,red] (sm2) -- (k4);
\node at (3.9,-0.2) (k5) {$V_{a_2}$};
\draw[-] (k5) -- (sm2);
\node at (-0.2,0.7) (k6) {$V_{a_1}$};
\node at (5.2,0.7) (k7) {$V_{a_3}$};
\draw[->] (k7) --  (0,3);
\draw[-] (k6) -- (sm2);
\end{tikzpicture}
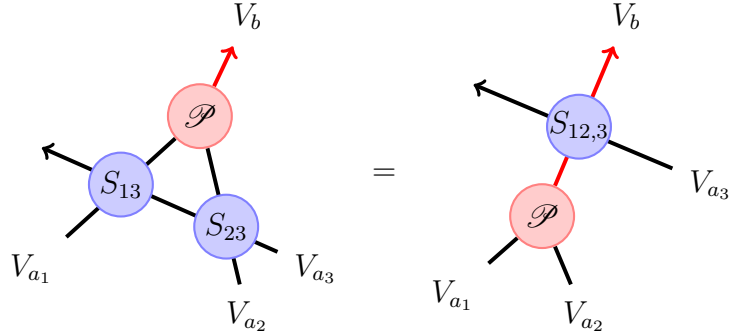
\caption{The bootstrap/fusion equations result from the equality of the diagrams above. One understands these diagrams in terms of localized wavepackets. The higher spin conserved charges implied by integrability can be used to move the trajectory of particle $a_3$ so that it either interacts with bound state $b$ or the particles $a_1$ and $a_2$ of which $b$ is composed.}
\label{f5}
}

The appropriate kinematical conditions are described in detail in \cite{Hoare:2012fc} but if particle 1 is coming in from the left then momenta are analytically continued as $p_1=\tilde p_1+ir$ and $p_2=\tilde p_2-ir$ with $r\in{\mathbb R}\geq0$. In terms of the auxiliary variables $x^\pm$, the condition that the bound state be on-shell leads to the possibilities shown in Figure \ref{f8}. Notice that it is more convenient here to label states with the pseudo rapidity $\nu$ rather than the rapidity $\theta$.
The processes I and II are pertinent to the magnons in the original theory. Processes III and IV are pertinent to the solitons in both the original and mirror theories. Note that the possible fusion rules for this S-matrix theory are
\EQ{
\text{I,}\ \text{III:}&\qquad b=a_1+a_2\ ,\\
\text{II,}\ \text{IV:}&\qquad b=|a_1-a_2|\ .
}
In the former I and III there is a constraint that $a_1+a_2\leq k$.
In the latter II and IV, Figure \ref{f8} shows the case $a_1>a_2$. What the rules in Figure \ref{f8} fail to specify is exactly on what sheet the bound state poles occur. This is described in detail in \cite{Hoare:2012fc}.

\FIG{
\begin{tikzpicture} [scale=0.8,line width=1.5pt,inner sep=2mm,
place/.style={circle,draw=blue!50,fill=blue!20,thick},proj/.style={circle,draw=red!50,fill=red!20,thick}]
\node at (1.8,5.7) (u1) {magnon};
\node at (9.8,5.7) (u1) {soliton};
\node at (-1.4,4.85) (v1) {II};
\draw (-1.7,-0.8) -- (5.2,-0.8) -- (5.2,5.2) -- (-1.7,5.2) -- (-1.7,-0.8);
\node at (2,2) [proj] (p1) {};
\node at (0.4,0.4) (i1) {$\langle a_1-1,0\rangle$};
\node at (0.4,-0.2) (l1) {$\nu-\frac{i\pi a_2}{2k}$};
\node at (3.6,0.4) (i2)  {$\langle a_2-1,0\rangle$};
\node at (3.6,-0.2) (l2) {$\nu+i\pi-\frac{i\pi a_1}{2k}$};
\node at (2,4.6) {$\langle a_1-a_2-1,0\rangle$};
\node at (2,4) (i3) {$(x_1^+,1/x_2^+)$};
\draw[->]  (i1) -- (p1);
\draw[->]  (i2) -- (p1);
\draw[<-,red]  (i3) -- (p1);
\node at (-0.1,2.2) {$\boxed{x_1^-=1/x_2^-}$};
\begin{scope}[xshift=8cm]
\node at (-1.3,4.85) (v1) {IV};
\draw (-1.7,-0.8) -- (5.2,-0.8) -- (5.2,5.2) -- (-1.7,5.2) -- (-1.7,-0.8);
\node at (2,2) [proj] (p1) {};
\node at (0.4,0.4) (i1) {$\langle 0,a_1-1\rangle$};
\node at (0.4,-0.2) (l1) {$\nu+\frac{i\pi a_2}{2k}$};
\node at (3.6,0.4) (i2)  {$\langle 0,a_2-1\rangle$};
\node at (3.6,-0.2) (l2) {$\nu-i\pi+\frac{i\pi a_1}{2k}$};
\node at (2,4.6) {$\langle 0,a_1-a_2-1\rangle$};
\node at (2,4) (i3) {$(1/x_2^-,x_1^-)$};
\draw[->]  (i1) -- (p1);
\draw[->]  (i2) -- (p1);
\draw[<-,red]  (i3) -- (p1);
\node at (-0.1,2.2) {$\boxed{x_1^+=1/x_2^+}$};
\end{scope}
\begin{scope}[yshift=7cm]
\node at (-1.4,4.85) (v1) {I};
\draw (-1.7,-0.8) -- (5.2,-0.8) -- (5.2,5.2) -- (-1.7,5.2) -- (-1.7,-0.8);
\node at (2,2) [proj] (p1) {};
\node at (0.4,0.4) (i1) {$\langle a_1-1,0\rangle$};
\node at (0.4,-0.2) (l1) {$\nu-\frac{i\pi a_2}{2k}$};
\node at (3.6,0.4) (i2)  {$\langle a_2-1,0\rangle$};
\node at (3.6,-0.2) (l2) {$\nu+\frac{i\pi a_1}{2k}$};
\node at (2,4.6) {$\langle a_1+a_2-1,0\rangle$};
\node at (2,4) (i3) {$(x_2^+,x_1^-)$};
\draw[->]  (i1) -- (p1);
\draw[->]  (i2) -- (p1);
\draw[<-,red]  (i3) -- (p1);
\node at (0.1,2.2) {$\boxed{x_1^+=x_2^-}$};\end{scope}
\begin{scope}[xshift=8cm,yshift=7cm]
\node at (-1.3,4.85) (v1) {III};
\draw (-1.7,-0.8) -- (5.2,-0.8) -- (5.2,5.2) -- (-1.7,5.2) -- (-1.7,-0.8);
\node at (2,2) [proj] (p1) {};
\node at (0.4,0.4) (i1) {$\langle 0,a_1-1\rangle$};
\node at (0.4,-0.2) (l1) {$\nu+\frac{i\pi a_2}{2k}$};
\node at (3.6,0.4) (i2)  {$\langle 0,a_2-1\rangle$};
\node at (3.6,-0.2) (l2) {$\nu-\frac{i\pi a_1}{2k}$};
\node at (2,4.6) {$\langle 0,a_1+a_2-1\rangle$};
\node at (2,4) (i3) {$(x_1^+,x_2^-)$};
\draw[->]  (i1) -- (p1);
\draw[->]  (i2) -- (p1);
\draw[<-,red]  (i3) -- (p1);
\node at (0.1,2.2) {$\boxed{x_1^-=x_2^+}$};\end{scope}
\end{tikzpicture}
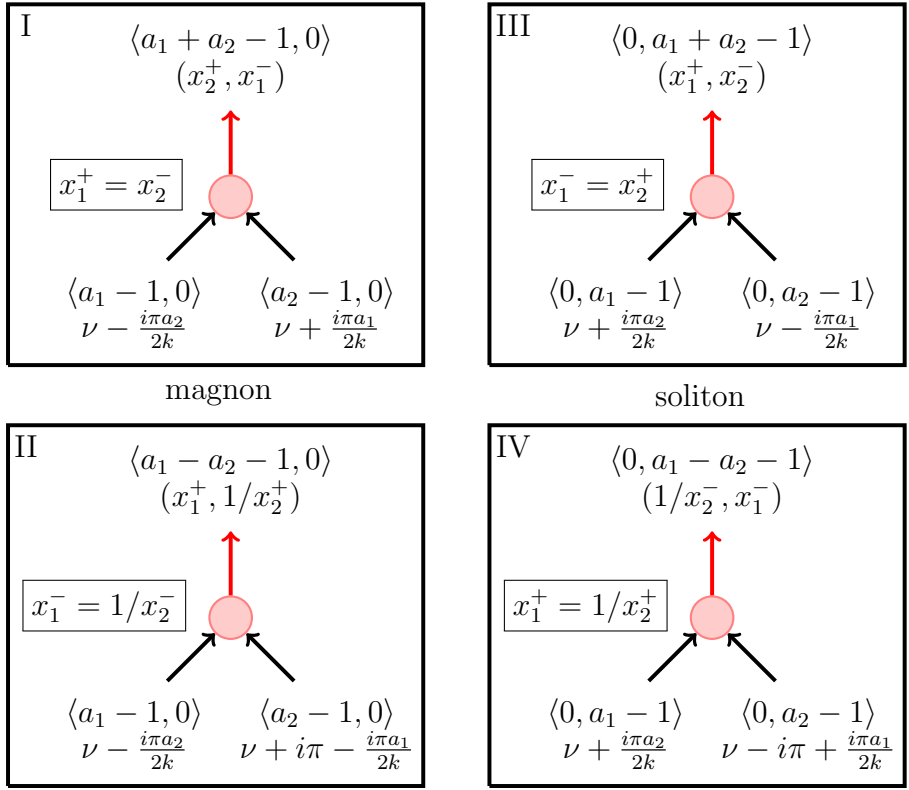
\caption{\small The possible bound state processes showing the representations and the incoming pseudo rapidities $\nu_1$ and $\nu_2$ in term of $\nu$, the pseudo rapidity of the bound state. Also shown is $(x_{12}^+,x_{12}^-)$ for the bound state. The processes II and IV have been written for the case $a_1>a_2$. The processes III and IV involving the solitons $\langle0,a-1\rangle$ are written for both the original and mirror theories but in the former after a shift of the rapidities by $i\pi/2$.}
\label{f8}
}

\noindent
{\bf Tests of the bootstrap}
\nopagebreak

The bootstrap is guaranteed to produce an S-matrix for particles transforming in the appropriate representations which lies in the commutant of the quantum group symmetry $U_q(\mh)$. This S-matrix should be the $R$-matrix of the quantum group appropriate to the representations under discussion up to an overall scalar factor: the so-called dressing phase. On the other hand the $R$-matrices for states transforming in the 
representations $\langle a-1,0\rangle$ or $\langle0,a-1\rangle$ have been deduced purely on symmetry grounds in \cite{deLeeuw:2011jr}. It is clearly important to compare these two methods for producing the bound-state S-matrix elements and in this section we turn to this problem.

The extent of our analysis will be quite modest, we will restrict ourselves to the scattering of the $a_1=2$ bound-state with the fundamental $a_2=1$ state for the magnon case. The essence of the test is based on the fact that the S-matrix has scalar sectors. This means that for a special choice of the incoming state $|\psi\rangle$ in the tensor product $V_{a_1}(\theta_1)\otimes V_{a_2}(\theta_2)$ it is mapped via the product of the S-matrix and the (graded) permutation $\mathbb P$ map to the same state up to a phase:
\EQ{
{\mathbb P}\cdot S(\theta_1,\theta_2)|\psi\rangle=e^{i\alpha}|\psi\rangle\ ,\qquad
|\psi\rangle\in V_{a_1}(\theta_1)\otimes V_{a_2}(\theta_2)\ .
} 
The S-matrix for the state $|\psi\rangle$ only involves a momentum dependent phase $\alpha(\theta_1,\theta_2)$ and is a scalar quantity.

We can see what this means very simply for the basic S-matrix $a_1=a_2=1$. In this case, there are scalar sectors corresponding to the states
\EQ{
|\phi^a(\theta_1)\rangle\otimes |\phi^a(\theta_2)\rangle\ ,\qquad
|\psi^a(\theta_1)\rangle\otimes |\psi^a(\theta_2)\rangle\ ,
}
(no sum over $a$). The ratio of the scalar S-matrix elements, or rather ${\mathbb P}\cdot S$, is then 
\EQ{
\frac A{-D}=\frac{U_1V_1}{U_2V_2}\cdot\frac{x_2^+-x_1^-}{x_2^--x_1^+}\ .
}
Obviously this ratio does not depend on the dressing phase and is solely dependent on the symmetry structure. 

For the case $a_1=2$ and $a_1=1$ there are again two distinct scalar sectors and therefore one way to test the two different constructions of the S-matrix is to compare the ratio of the two scalar amplitudes. First of all, we describe how this ratio follows from the symmetry analysis in \cite{deLeeuw:2011jr}. Using the notation of \cite{deLeeuw:2011jr}, the two distinct scalar sectors correspond to the states $|0\rangle\equiv|0,0,0,2\rangle\otimes |0,0,0,1\rangle$ and $|0,0\rangle^I\equiv|0,1,0,1\rangle\otimes|0,1,0,0\rangle$ in the tensor product $V_2\otimes V_1$. The relation between these states and the states in our notation is as follows. Firstly for the states in the basic representation $V_1$,
\EQ{
|0,0,0,1;\theta\rangle\thicksim|\phi^1(\theta)\rangle\ ,
\qquad
|0,1,0,0;\theta\rangle\thicksim|\psi^1(\theta)\rangle\ .
}
Then for the states in the bound state $\langle1,0\rangle$, which we realize in terms of 2 basic particle states evaluated at the rapidities shown in I of Figure \ref{f8}, we have
\EQ{
|0,0,0,2;\theta\rangle&\thicksim|\phi^1(\theta-i\pi/2k)\phi^1(\theta+i\pi/2k)\rangle\ ,\\
|0,1,0,1;\theta\rangle&\thicksim \gamma_1\sqrt{x_2^+}|\psi^1(\theta-i\pi/2k)\phi^1(\theta+i\pi/2k)\rangle\\ &+\gamma_2\sqrt{qx_1^+}|\phi^1(\theta-i\pi/2k)\psi^1(\theta+i\pi/2k)\rangle\ ,
}
up to an overall normalization.

The matrix $\mathbb S$ of \cite{deLeeuw:2011jr} is our $\mathbb P\cdot S$ normalized so that the element in the scalar sector for the state $|0\rangle$ is $1$. The ratio of the two scalar elements of $\mathbb S$ for $|0,0\rangle^I$ relative to $|0\rangle$ is then what is defined in \cite{deLeeuw:2011jr} to be\footnote{In the notation of \cite{deLeeuw:2011jr}, $a_1=M_1=2$ and $a_2=M_2=1$.}
\EQ{
\mathscr D=q^{1/2}\frac{U_2V_2}{U_1V_1}\frac{x_1^+-x_2^-}{x_1^--x_2^+}\ .
}
By brute force, we find precisely the result above from the bootstrap equation.

\section{Representations of \texorpdfstring{$U_q(\msu(2))$}{Uq(su(2))}}\label{a5}

Here we review  {the main} features of the representation theory of $U_q(\msu(2))$ when the deformation parameter is a root of unity. 
We will mostly reproduce some parts of the review article~\cite{Arnaudon:1992ar}. More extensive information about quantum groups and their representations and references to the original literature can be found in~\cite{Chari:1994pz,Gomez:1996az}.
The realization of these representations in terms of $q$-oscillators has been discussed in~\cite{Arutyunov:2012zt}.

The quantum group $U_q(\msu(2))$ is generated by $J_+$, $J_-$ and $K$ which satisfy the defining relations
\EQ{
KJ_\pm K^{-1}= q^{\pm2}J_\pm\,,\qquad
[J_+,J_-]= \frac{K-K^{-1}}{q-q^{-1}}\,,\qquad KK^{-1}=K^{-1}K=1
\,.
\label{cr}
}
They commute with the $q$-deformed quadratic Casimir
\EQ{
C= J_- J_+ +  \frac{qK +q^{-1}K^{-1}}{(q-q^{-1})^2}\,.
\label{casimir}
}
$U_q(\msu(2))$ is a particular deformation of the universal enveloping algebra of the Lie algebra $\msu(2)$, which can be made explicit by writing $K=q^H$ so that the $q\to1$ limit of~\eqref{cr} becomes simply
\EQ{
[H,J_\pm]= \pm2 J_\pm\,,\qquad 
[J_+,J_-]= 2H\,.
}
However, it is worth noticing that not all the representations of $U_q(\msu(2))$ defined in terms of $J_\pm,K$ make sense in terms of $J_\pm,H$ (for example, see~\cite{Chari:1994pz}, Chapter~10, and below).
$U_q(\msu(2))$ has an associated co-product $\Delta$ which describes how the generators act on tensor products. With the conventions of~\cite{Arnaudon:1992ar}, it is written as
\EQ{
\Delta(K)=K\otimes K,\quad
\Delta(J_+)=J_+\otimes 1+ K\otimes J_+,\quad \Delta(J_-)=J_-\otimes K^{-1}+ 1\otimes J_-.
\label{cpA}
}

\noindent{\bf E.1\quad Representations for $q$~not being a root of unity~\cite{Lusztig:1988at,Rosso:1988gg}}
\nopagebreak

\noindent In this case the representations of $U_q(\msu(2))$
are very similar to those of $\msu(2)$. The
finite dimensional irreducible representations $V_j^{(\sigma)}$ are labelled by a half-integer spin $j=0,\frac12,1,\ldots$ and a discrete parameter $\sigma=\pm1$. They have dimension $2j+1$ and a basis $\{\omega_0,\omega_1,\ldots,\omega_{2j}\}$ on which the action of the generators is
\EQ{
&K\omega_p = \sigma\, q^{2j-2p} \omega_p\,,\\[5pt]
&J_-\omega_p= \omega_{p+1}\,,\quad p=0,\ldots, 2j-1\,,\qquad J_-\omega_{2j}=0\,,\\[5pt]
&J_+\omega_p= \sigma\, [p][2j-p+1]\, \omega_{p-1}\,,\quad p=1,\ldots, 2j\,,\qquad J_+\omega_{0}=0\,,
\label{repA}
}
where we have used the $q$-number
\EQ{
[n]=\frac{q^n-q^{-n}}{q-q^{-1}}\ .
}
If $K=q^H$ and  $\sigma=+1$, note that the representation $V_j^{(+1)}$ can be expressed directly in terms of $H$:
\EQ{
H\omega_p = 2(j-p) \omega_p\,.
\label{noH}
}
However, this is not possible for the representations with $\sigma=-1$ (see~\cite{Chari:1994pz}, Chapter~10).

\noindent
{\bf E.2\quad Representations when $q$~is a root of unity~\cite{Roche:1988te,DeConcini:1990}}
\nopagebreak

\noindent
Let $m'$ be the smallest integer such that $q^{m'}=1$ and define
\EQ{
m=\begin{cases}
 m' & \text{if $m'$ is odd,} \\
 \frac{m'}2     & \text{if $m'$ is even.}
\end{cases}
}
In our case $q=e^{i\p/k}$ and, thus, $m=k$. 
Compared to the case when $q$ is not a root of unity, the main difference is the structure of the centre of $U_q(\msu(2))$ which now is larger. Namely, in addition to the $q$-deformed quadratic Casimir $C$, it contains also $J_+^{m}$, $J_-^{m}$, and $K^{m}$. We will denote by $x$, $y$, $z$ and $c$ the values of $J_+^{m}$, $J_-^{m}$, $K^{m}$, and $C$ on finite dimensional irreducible representations.
Then, their dimension is bounded by $m$, and the irreducible representations of dimension $m$ depend on three complex continuous parameters. In the following, we will call type A irreducible representations those that have a classical ($q\to1$) analogue and type B the others.

\noindent
{\bf Type A representations:}
the irreducible representations of type~A are labelled by a half-integer spin $j$ such that $0\leq 2j+1\leq m$ and a discrete parameter $\sigma=\pm1$. They have a basis $\{\omega_0,\omega_1,\ldots,\omega_{2j}\}$ on which the action of the generators is given by~\eqref{repA}. Therefore, we will use the same notation $V_j^{(\sigma)}$ to denote them. On these representations the central elements take the values
\EQ{
x=y=0\,,\qquad z= (\sigma q^{2j})^m=\pm1\,,\qquad c=\sigma\,\frac{q^{2j+1}+q^{-(2j+1)}}{(q-q^{-1})^2}\,.
}
In particular, notice that $J_\pm^m=0$.

\noindent
{\bf Type B representations:}
these representations are characterized by three complex parameters $\beta$, $y$ and $\lambda=q^\mu$. They have dimension~$m$ and one can choose a basis $\{v_0,v_1,\ldots,v_{m-1}\}$ such that
\EQ{
&K v_p= \lambda q^{-2p} v_p\,,\\[5pt]
&J_-v_p = v_{p+1} \,,\quad p=0,\ldots, m-2\,,\qquad J_-v_{m-1}= yv_0\,, \\[5pt]
&J_+v_p= \left([p][\mu-p+1]+y\beta\right) v_{p-1}\,,\quad p=1,\ldots,m-1\,,\qquad J_+v_0= \beta v_{m-1}\,.
}
The central elements $J_+^m$, $J_-^m$,$K^m$, and C take the values
\EQ{
x=\beta\prod_{p=1}^{m-1} \left([p][\mu-p+1]+y\beta\right)\,,
}
$y$, $z=\lambda^m$, and $c=y\beta +(q-q^{-1})^{-2}(q\lambda +q^{-1}\lambda^{-1})$, respectively. 
This representation can be denoted either by $B(x,y,z,c)$ or $B'(\beta,y,\lambda)$, and notice that $B'(0,0, \pm q^{m-1})=V_{\frac{m-1}2}^{(\pm1)}$. It is irreducible if one of the three following conditions is satisfied:
\EQ{
&a)\quad x\not=0\,,\hspace{11cm}\,\\[5pt]
&b)\quad y\not=0\,,\\[5pt]
&c)\quad \beta=0\; \text{and}\; \lambda^2 \in{\mathbb C}\backslash\{1,q^2,\ldots,q^{2(m-2)}\} \,.\\[5pt]
}
The representation $B(x,y,z,c)$ is called {\em cyclic} if $xy\not=0$. In this case it has neither highest-weight nor lowest-weight vectors. The representation is called {\em semi-cyclic} if either $x=0\not=y$ or $y=0\not=x$, and it has only either a highest-weight or a lowest-weight vector. Finally, it is called {\em nilpotent} if $x=y=0$ and $\lambda$ is generic, in which case if has both a highest-weight and a lowest-weight vector.

\noindent
{\bf E.3\quad Tensor product of type~A representations~\cite{Pasquier:1989kd,Keller:1991}}
\nopagebreak

\noindent
For the S-matrix theories discussed in Section~\ref{s2}, the basis states  transform in the spin-$\frac12$ representation $V_{\frac12}\equiv V_{\frac12}^{(+1)}$ and multi-particle states transform in the tensor product representation ${V_{\frac12}}^{\otimes N}$. When $q$ is not a root of unity these tensor products decompose into irreducible representations $V_j^{(+1)}\equiv V_j$ with $j\leq \tfrac{N}2$. However, when $q$ is a root of unity the decomposition becomes more complicated and involves both irreducible and  {reducible but indecomposable} representations. For our purposes it will be enough to summarize the decomposition of tensor products of representations of type~A only.

The tensor product of two (type~A) representations $V_{j_1}^{(\sigma_1)}$ and $V_{j_2}^{(\sigma_2)}$ decomposes into irreducible representations of the same type and, if $2(j_1+j_2)+1>m$, into some indecomposable spin representations. The relevant indecomposable spin representations $\text{Ind\,}(j,\sigma)$ have dimension $2m$ and are labelled by a half-integer spin $j$ such that $1\leq 2j+1<m$ and a discrete parameter $\sigma=\pm1$. They have a basis $\{\omega_0,\ldots, \omega_{m-1}, x_0,\ldots,x_{m-1}\}$ on which the action of the generators is
\EQ{
&K\omega_p=\sigma q^{-2j-2-2p} \omega_p\,,\\[5pt]
&J_-\omega_p= \omega_{p+1}\,,\quad p=0,\ldots,m-2\,,\qquad J_-\omega_{m-1}=0\,,\\[5pt]
&J_+\omega_p=\sigma[p][-2j-p-1]\omega_{p-1}\,,\quad 
p=0,\ldots,m-1\,,\\[5pt]
&Kx_p=\sigma q^{2j-2p} x_p\,,\\[5pt]
&J_-x_p= \omega_{p+1}\,,\quad p=0,\ldots,m-2\,,\qquad J_-x_{m-1}=0\,,\\[5pt]
&J_+x_p= J_-^{p+m-2j-2}\omega_0 + \sigma[p][2j-p+1]x_{p-1}\,,\quad p=0,\ldots,m-1\,.
}
In particular, $J_+x_0=\omega_{m-2j-2}$, $J_+x_{2j+1}=\omega_{m-1}$, and $J_\pm^m=0$.
This indecomposable representation contains the sub-representation $V_j^{(\sigma)}$.

The decomposition of the tensor product of two irreducible representations of type~A is
\EQ{
V_{j_1}^{(\sigma_1)}\otimes V_{j_2}^{(\sigma_2)}= \left[\bigoplus_{j=|j_1-j_2|}^{\text{min}(j_1+j_2, m-j_1-j_2-2)}\, V_{j}^{(\sigma_1\sigma_2)}\right]
\bigoplus \left[\bigoplus_{j=m-j_1-j_2-1}^{(m-1)/2} \, \text{Ind\,}(j,\sigma_1\sigma_2) \right]\,,
}
where the sums are restricted to integer values of $j$ if $j_1+j_2$ is integer and to half-integer values if $j_1+j_2$ is half-integer, and $\text{Ind\,}(\frac{m-1}2,\sigma)\equiv V_{\frac{m-1}2}^{(\sigma)}$. Notice that in the first sum $j$ is always bounded to be $\leq\frac{m}2-1$. The decomposition of tensor products of type~A representations and indecomposable spin representations is schematically given by
\EQ{
&V_{j_1}^{(\sigma_1)}\otimes \text{Ind\,}(j_2,\sigma_2)= \bigoplus_j\, \text{Ind\,}(j,\sigma_1\sigma_2)\,,\\[5pt]
&\text{Ind\,}(j_1,\sigma_1)\otimes \text{Ind\,}(j_2,\sigma_2)= \bigoplus_j\, \text{Ind\,}(j,\sigma_1\sigma_2)\,.
}
All this makes it possible to restrict the set of allowed representations to $V_j^{(\sigma)}$ with \mbox{$j\leq \tfrac{m}2-1$} and introduce a {\em truncated} tensor product where the indecomposable representations do not appear
\EQ{
V_{j_1}^{(\sigma_1)}\, \widetilde{\otimes}\, V_{j_2}^{(\sigma_2)}= \bigoplus_{j=|j_1-j_2|}^{\text{min}(j_1+j_2, m-j_1-j_2-2)}\, V_{j}^{(\sigma_1\sigma_2)}\,,
\label{ttp}
}
a truncation that is well known in the context of conformal field theory (for example, see~\cite{Gomez:1996az}). Notice also that the truncated tensor product can be naturally restricted to the representations with $\sigma=+1$, which are the only ones that are relevant for our purposes. The truncated decomposition is explicitly described in terms of the $q$-deformed Clebsch-Gordan coefficients defined by~\eqref{qCGor} whose value is not modified by the truncation~\cite{Ardonne:2010zu}.

\end{document}